\RequirePackage{fix-cm}
\documentclass[twocolumn,epjc3]{svjour3}  
\smartqed  
\RequirePackage{graphicx}
\RequirePackage{latexsym}
\RequirePackage[numbers,sort&compress]{natbib}
\RequirePackage[colorlinks,citecolor=blue,urlcolor=blue,linkcolor=blue]{hyperref}
\usepackage{url}
\usepackage{mathtools}
\usepackage[toc,page]{appendix}
\newcommand*{\Scale}[2][4]{\scalebox{#1}{$#2$}} 
\journalname{Eur. Phys. J. C}
\begin{document}

\title{Viability tests of \textit{f(R)}-gravity models with Supernovae Type 1A data}


\author{R.T. Hough \thanksref{e1,addr1}
        \and
        A. Abebe \thanksref{addr2} 
        \and
        S.E.S. Ferreira \thanksref{addr1}
}

\thankstext{e1}{e-mail: renierht@gmail.com}


\institute{Centre for Space Research, North-West University, Potchefstroom 2520, South Africa \label{addr1}
           \and
           Centre for Space Research, North-West University, Mahikeng 2745, South Africa \label{addr2}
}

\date{Received: date / Accepted: date}

\maketitle

\begin{abstract}
In this work, we will be testing four different general \textit{f(R)}-gravity models, two of which are  the more realistic models (namely the Starobinsky and the Hu-Sawicki models), to determine if they are viable alternative models to pursue a more vigorous constraining test upon them. For the testing of these models, we use 359 low- and intermediate-redshift Supernovae Type 1A data obtained from thRede SDSS-II/SNLS2 Joint Light-curve Analysis (JLA). We develop a Markov Chain Monte Carlo (MCMC) simulation to find a best-fitting function within reasonable ranges for each \textit{f(R)}-gravity model, as well as for the Lambda Cold Dark Matter ($\Lambda$CDM) model. For simplicity, we assume a flat universe with a negligible radiation density distribution. Therefore, the only difference between the accepted $\Lambda$CDM model and the \textit{f(R)}-gravity models will be the dark energy term and the arbitrary free parameters. By doing a statistical analysis and using the $\Lambda$CDM model as our ``true model", we can obtain an indication whether or not a certain \textit{f(R)}-gravity model shows promise and requires a more in-depth view in future studies. In our results, we found that the Starobinsky model obtained a larger likelihood function value than the $\Lambda$CDM model, while still obtaining the cosmological parameters to be $\Omega_{m} = 0.268^{+0.027}_{-0.024}$ for the matter density distribution and $\bar{h} = 0.690^{+0.005}_{-0.005}$ for the Hubble uncertainty parameter. We also found a reduced Starobinsky model that are able to explain the data, as well as being statistically significant.

\end{abstract}

\section{Introduction}\label{intro}
Since the proposition of the Theory of General Relativity (GR) by Einstein, it has developed into the accepted theory to explain gravity. What made GR useful was that it was not only able to explain extreme gravity phenomena, but was also able to reduce back to a Newtonian description of gravity in a weak gravitational field. Due to the ability of GR to explain the expansion of the Universe \cite{Kirshner2004}, the Hot Big Bang theory was developed using GR as its mathematical basis. However, in recent times, it was discovered that the expansion of the Universe was accelerating \cite{Nojiri2017}, which is not in line with the GR predictions. Therefore, an unknown pressure force acting out against gravity, called ``dark energy"  ($\Lambda$ $\sim$ cosmological constant) was added to explain why gravity on cosmological scales were not able to slow down the expansion \cite{Capozziello2018}. By using the Einstein-Hilbert action, which tries to extremize the path between two time-like points in spacetime, with the inclusion of dark energy, one can derive the cosmological field equations. From the cosmological field equation, the Friedmann equations can be derived, with these equations being able to explain the accelerated expansion of the Universe in the Big Bang model, and is given by \cite{Abebe2013, Romeu2014}
\begin{eqnarray}
 H^{2}(t) &=& \frac{\rho (t)}{3} - \frac{\kappa}{a^{2}(t)}+\frac{\Lambda}{3},\\
 \dot{H}(t) &=& -H(t)^{2} -\frac{1}{6}\Big[\rho (t)+3P(t)\Big]+\frac{\Lambda}{3},
\end{eqnarray}
where $H(t) = \frac{\dot{a}(t)}{a(t)}$ is the Hubble parameter with $a(t)$ the scale factor describing the relative size of the Universe at a certain time, $\rho (t)$ is the energy density, $P(t)$ is the isotropic pressure, and $\kappa$ is the 3D (spacial) curvature. Furthermore, to derive these particular Friedmann equations, we had to assume we have a Friedmann-Lema\^{\i}tre-Robertson-Walker (FLRW) spacetime metric, as well as normalizing the system by using a geometric unit description where $c=1=8\pi G$. To close the Friedmann equation system, we had to use the equation of state parameter $\omega$, by relating $\rho$ and $P$. We also assumed a perfect fluid, therefore $\omega$ is constant \cite{Trodden2004}\footnote{Even though we will be using a constant $\omega$, there exist studies that tries to parametrise the equation of state for both the $\Lambda$CDM model and the modified gravity models \cite{Planck2018, Jaime2018}.}. This closed system is called the Lambda Cold Dark Matter ($\Lambda$CDM) model. 

However, since dark energy is an unknown pressure force, this poses a problem: What is dark energy and why does it account for the majority content of the Universe ($\sim 68\%$) \cite{Planck2018}? Other questions that also arises within the $\Lambda$CDM model, is the Horizon and Flatness problems that stem from an early-time accelerated expansion epoch, called the Inflation epoch \cite{Felice2010, Starobinsky1980, Guth1981, Capozziello2019a}. Other known problems in the $\Lambda$CDM model are the magnetic monopole problem and the matter/anti-matter ratio problem \cite{Capozziello2019a, Vieira2011}. Due to these problems, it has been previously suggested that we need to modify our theory of gravity. One such theory is \textit{f(R)}-gravity. This theory makes the modification within the Einstein-Hilbert action by changing the Ricci scalar $R$ to a generic function dependent on ($R$), therefore replacing the dark energy term with arbitrary free parameters. Re-deriving the Friedmann equations with this modification, we obtain \cite{ Abebe2013, Felice2010, Romero2018}
\begin{eqnarray}\label{eq: f(R) Friedmann equation}
	H^{2}(t) &=&\frac{\rho (t)}{3f^{\prime}} - \frac{\kappa}{a^{2}(t)}+\frac{1}{6}\bigg[R - \frac{f}{f^{\prime}}\bigg]- H\dot{R}\frac{f^{\prime\prime}}{f^{\prime}},\\
	\dot{H}(t) &=& -H^{2}(t)- \frac{\rho(t)}{3f^{\prime}}+\frac{f}{6f^{\prime}}+H\dot{R}\frac{f^{\prime\prime}}{f^{\prime}},
\end{eqnarray}
where $f = f(R)$, with $f^{\prime}$ and $f^{\prime\prime}$ being the first- and second- derivatives of the generic function w.r.t. $R$.
\section{Supernovae cosmology and MCMC simulations}\label{sec:Supernovae cosmology}
\subsection{Supernovae cosmology}\label{sec: distance modulus}
To test Eq. \ref{eq: f(R) Friedmann equation}, we will use Supernovae Type 1A data. This class of supernovae is the resultant of a white dwarf (WD) star accreting a low-mass companion star until the accreted Hydrogen outer-layer, from the companion star, is compressed to the point that the WD explodes \cite{Tayler1994}. Since this process is always the same, their luminosities are relatively similar and can therefore be regarded as standard candles \cite{Barbon1973, Richardson2002}. Their measured flux is therefore only dependent on the distance to the particular supernova. We will use redshift ($z$) to approximate the distance. This will allow us to use the distance modulus function to test the expansion of the Universe, since the distance to the supernovae is changing. For simplicity, we will assume a flat universe ($\Omega_{k}=0$), with a negligible radiation density ($\Omega_{r}\approx 0$). The distance modulus function we obtain, by using the combination of different distance definitions found in \cite{Deza2009}, is given as\footnote{We will be using the distance modulus function in terms of $Mpc$.}
\begin{equation}
	\mu = m-M = 25+5\times \log_{10}\Bigg(3000\bar{h}^{-1}(1+z)\int ^{z}_{0} \frac{dz^{\prime}}{h(z^{\prime})}\Bigg),
\label{eq: distance modulus}
\end{equation}
where $m$ is the apparent magnitude and $M$ is the absolute magnitude of the measured supernova, while per definition we have the Hubble uncertainty parameter as $\bar{h} = \frac{H(z)}{100\frac{km}{s.Mpc}}$ with $H(z)$ is the Hubble parameter. Now that we have a model, we will use 359 low- and intermediate redshift supernovae data obtained from the SDSS-II/SNLS3 Joint Light-curve Analysis (JLA). However, using only Supernovae Type 1A data means that we will not be able to fully constrain the Hubble uncertainty parameter, due to $H_{0}$ being degenerate with the absolute magnitude $M$ of the particular supernovae \cite{Colg2019}. However, since the absolute magnitude is close to being constant and not expected to dependent on redshift, its combination with the $H_{0}$ still gives a robust enough prediction that we can get an idea on the validity of the different models. To break this degeneracy, Cepheid variable star data will need to be incorporated \cite{Riess2009, Riess2011} in future studies, where more accurate predictions, on the potential viable models that we find in this study, can be made by incorporating the different data sets on a state-of-the-art software program as mentioned in sec. \ref{sec: results}

The reason for using low- and intermediate redshift data is to have within our data the transition phase between the decelerated expansion (matter dominated) epoch and the late-time acceleration (dark energy) epoch which only started at around $z\approx 0.5$ \cite{Linder2001, Capozziello2019}. We will be using the absolute magnitudes for these supernovae in the B-filter that can be obtained from the research papers \cite{Conley2010, Neill2009, Hicken2009}\footnote{These absolute magnitudes can also be found on NASA's Extragalactic Database (NED).}. This method is called supernovae cosmology.

\subsection{Markov Chain Monte Carlo (MCMC) simulation}
To find the best-fitting distance modulus for each model, we will use MCMC simulations. These simulations are able to search for the most probable free parameter value, given certain physical constrains. In particular, we will be using the Metropolis-Hastings (MH) algorithm \cite{Roberts1994, Chib2001}, which starts by calculating the likelihood for each initial chosen free parameter's distance modulus. The simulation then takes a random step for each parameter away from the initial conditions, but within the physical constrains. It then calculates the likelihood for each possible combination between the initial conditions and the random parameter values, to find the combination that has the largest likelihood of occurring. 

The simulation then finds an acceptance ratio between the initial condition likelihood  and the new largest likelihood combination. If the new combination has a acceptance ratio value large than 1, it is accepted. If it is lower than 1, a chance is created for the second combination to still be accepted in ratio to the probability for each combination to occur. After the acceptance or rejection of a certain combination, the algorithm starts at the top again. Since, we need a probability distribution to be able to calculated the likelihood for each parameter value's distance modulus, we assume, for simplicity, a Gaussian distribution.

We use the \textit{EMCEE Hammer Python} package to execute the MCMC simulation. This package uses different random walkers (in most cases we will use 100), each executing the MH algorithm and all starting at the same initial parameter values and converging on the most probable parameter values. The last iteration then creates a Gaussian distribution based on each random walker's ending parameter values. Using the average values for each probability distribution for each parameter, we will have on average the best-fitting parameter value and its 1$\sigma$-deviation for each free parameter.

\subsection{AIC and BIC statistical analysis}\label{sec:stats}
To test whether or not these \textit{f(R)}-gravity models are able to explain the data, we will use the Akaike Information Criterion (AIC) and the Bayesian/Schwarz Information Criterion (BIC) selection methods \cite{Burnham2004}. These selection criteria uses the likelihood function value of each of the best-fitting models, while taking into account the amount of free parameters the model use. This is important, since a model that uses more free parameters can fit the data more precisely (has more freedom to change the shape of the function), but might not be as valuable as another model that uses less free parameters.  The AIC and BIC selections are given as 

\begin{eqnarray}\label{eq: AIC and BIC criteria}
	\bullet\quad AIC &=& \chi ^{2} +2K,\\
	\bullet\quad BIC &=& \chi ^{2} +K\log (n),
\end{eqnarray}
where $\chi^{2}$ is calculated by using the model's Gaussian likelihood function $\mathcal{L}(\hat{\theta} |data)$ value, $K$ is the amount of free parameters for the particular model, while $n$ is the amount of data points in our dataset. Since the AIC and BIC selection values can be any positive value, we need to compare the particular \textit{f(R)}-gravity model's AIC and BIC values to that of a ``true model" (in this case the $\Lambda$CDM model) \cite{Nunes2017}, by finding the difference between them. We will be using the Jeffrey's scale in order to make conclusions about the \textit{f(R)} model. It should be noted that this scale is not exclusive and should be handled with care \cite{Nesseris2013}. The Jeffreys scale ranges are:
\begin{eqnarray}\label{eq: Jeffreys scale}
	&\bullet &\quad \Delta IC \leq 2 - \textrm{substantial support},\\
	&\bullet &\quad 4\leq \Delta IC \leq 7 - \textrm{less support},\\
	&\bullet &\quad \Delta IC > 10 - \textrm{no observational support}.
\end{eqnarray}

\section{Results}\label{sec: results}
\subsection{The $\Lambda$CDM model}
We will use the $\Lambda$CDM model to calibrate our MCMC simulation, as well as use it as our ``true model" to which we can compare the \textit{f(R)}-gravity models against to find if they are viable alternative models. By assuming a flat universe with negligible radiation density, we can find a normalized Friedmann equation for the $\Lambda$CDM model in terms of redshift, with the substitution $\Omega_{\Lambda} =1-\Omega_{m}$ \cite{Bennett2014, Odintsov2017}, as
\begin{equation}
	h(z) = \sqrt{\Omega_{m}\big(1+z\big)^{3}+1-\Omega_{m}}.
\label{eq. normalized friedmann equation}
\end{equation}
To execute the MCMC simulation for the $\Lambda$CDM model, we need to combine eq. \ref{eq. normalized friedmann equation} with eq. \ref{eq: distance modulus}. The MCMC simulation gave the cosmological parameter values for the $\Lambda$CDM model, based on our test supernovae dataset, as $\Omega_{m} = 0.268^{+0.025}_{-0.024}$ for the matter density distribution and $\bar{h} =  0.697^{+0.005}_{-0.005}$ for the Hubble uncertainty parameter. These values are in line with other Supernovae Type 1A cosmological results, even though they are not within $1\sigma$ from the Planck2018 results that were determined on the Cosmic Microwave Background (CMB) radiation data. This discrepancy between early-time data, such as the CMB, and the late-time data, such as the supernovae events, have been shown to exist \cite{Riess2016, Capozziello2019a}. Therefore, after finding possible viable \textit{f(R)}-gravity model using only the one dataset, we must continue in testing those potential models on different datasets for a more comprehensive in-depth study for constraining these alternative models. 

This discrepancy is not only limited to these two methods of calculating the cosmological parameter values. In a paper by \cite{Mortsell2018}, they showed that different experiments resulted in different $H_{0}$ values. With all the local measurements, such as eclipsing binaries in the Large Magellanic Clouds or Cepheid stars within the Milky way, tend to result in higher values for the Hubble constant, while the early-time data tended to give a lower Hubble constant value. In future work, we can combine Supernovae Type 1A with CMB data to be able to show this discrepancy. It will also be worth it to test our potential viable \textit{f(R)}-gravity models on other datasets, such as $H(z)$ and BAO \cite{cao2018}, to see how the different \textit{f(R)}-gravity model lead to different contributions from the matter and dark energy densities distributions within the Universe \cite{Odintsov2017, Odintsov2018}. 

\begin{figure}[tbp]
\centering 
\includegraphics[width=.48\textwidth,trim = 20 5 45 20,clip]{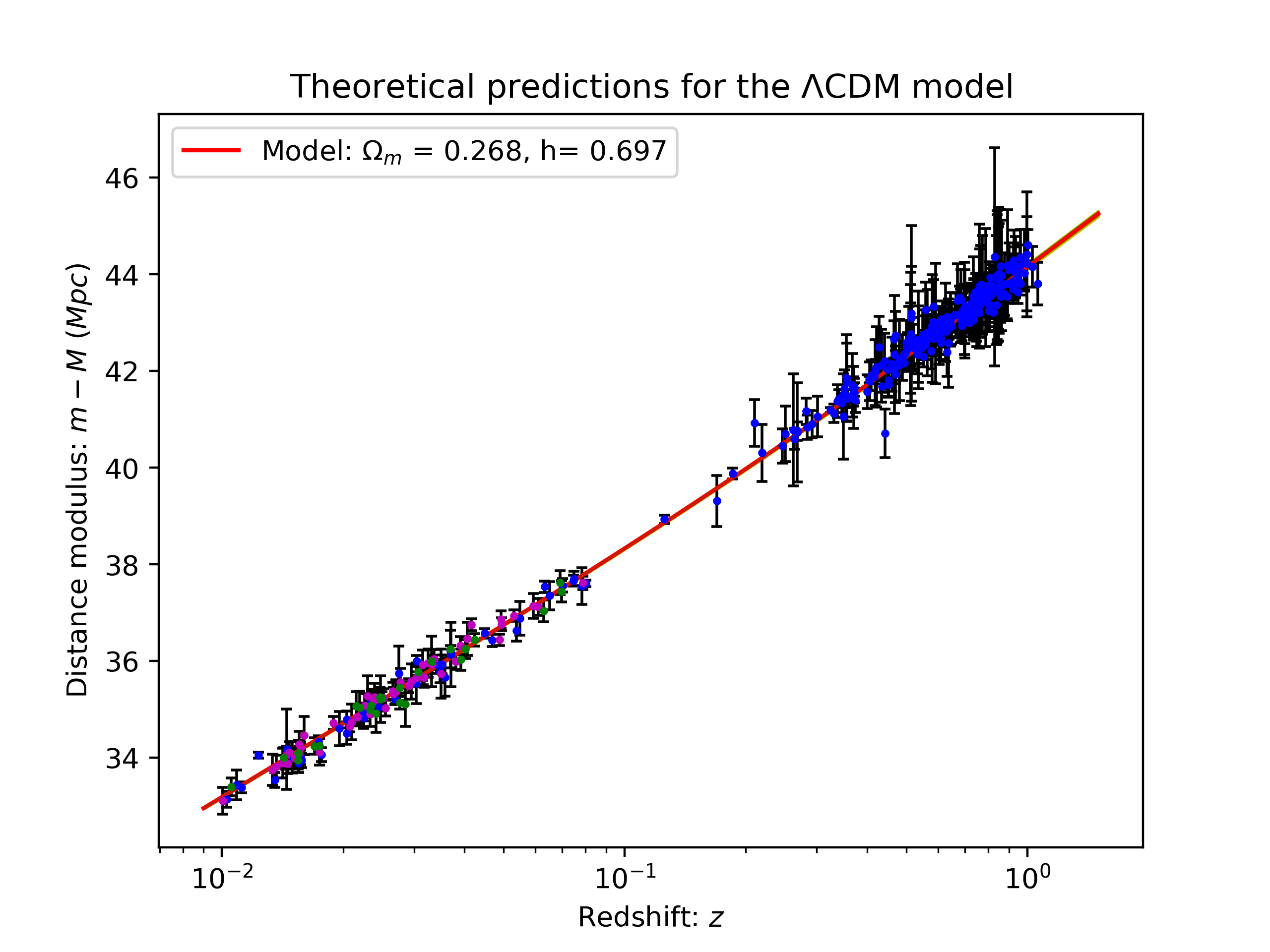}
\hfill
\includegraphics[width=.48\textwidth,trim = 20 5 45 20,clip]{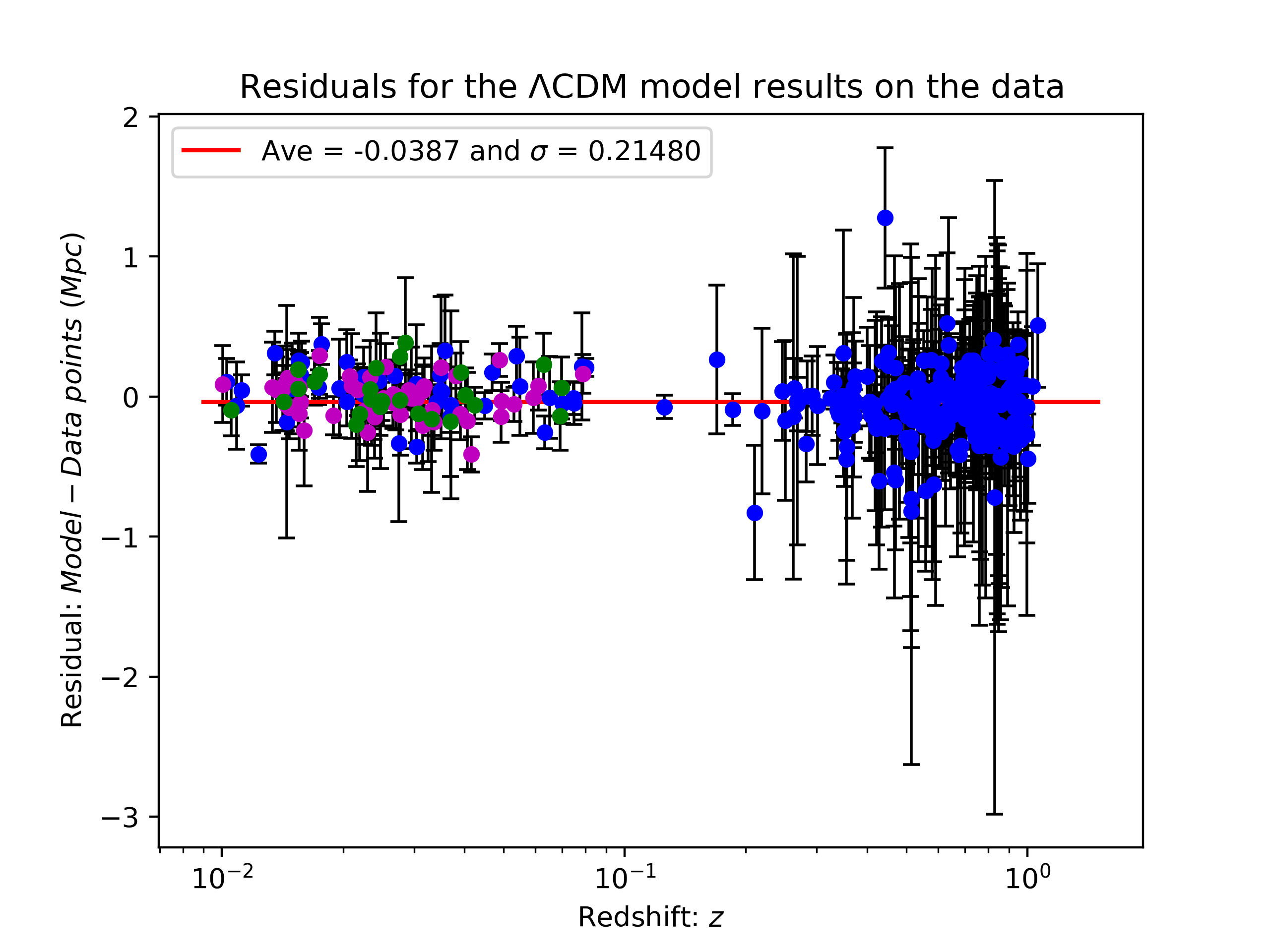}
\caption{The $\Lambda$CDM model's best-fit to the Supernovae Type 1A data (top panel), with the cosmological parameter values as $\Omega_{m} = 0.268^{+0.025}_{-0.024}$ (constrained) and $\bar{h} =  0.697^{+0.005}_{-0.005}$, respectively. Furthermore, the residuals between the model's predicted distance modulus values and the actual data points are also shown (bottom panel).}
\label{fig: LCDM model result graphs}
\end{figure}

Now that we have discussed the MCMC results and have shown that the results are in line with expectation, we can make a plot for the best-fitting $\Lambda$CDM model on the Supernovae Type 1A data, to which we can the compare the \textit{f(R)}-gravity models to. This result is shown in figure \ref{fig: LCDM model result graphs}.

From figure \ref{fig: LCDM model result graphs}, we can also confirm that the MCMC simulation's calibration was done correctly, since the $\Lambda$CDM model fits the data with quite a high accuracy, as well as not having an over- or under-estimation at various different redshifts. As a note for the rest of the models, the average residual value that is shown on the residuals graphs shows the average amount the model over- or under-estimates the distance modulus ($Mpc$) for each supernovae. Therefore, the $\Lambda$CDM under-estimates the supernovae distance modulus, on average, with $\bar{x}_{res} = -0.0387$ Mpc, and the standard deviation of the data on this average distance is $\sigma_{res} =0.21480$, showing that this is a very tight relation. Furthermore, in terms of constraining the parameters, the MCMC simulation were able to constrain both the cosmological parameters.

\subsection{\textit{f(R)}-gravity model results}
We can now advance to the testing of various \textit{f(R)}-gravity models. We will use two toy models, namely $f(R) = \beta R^{n}$ and $f(R) = \alpha R+\beta R^{n}$ \cite{Abebe2013}, as well as two realistic models, namely the Starobinsky and Hu-Sawicki models, which are given by \cite{Cardone2012, Starobinsky2007, Tsujikawa2008b}

\begin{eqnarray}\label{eq: Starobinsky and Hu-Sawicki models}
	f(R) &=& R +\beta R_{c}\Bigg[\bigg(1+\frac{R^{2}}{R_{c}^{2}}\bigg)^{-n}-1\Bigg],\\
	f(R) &=& R -\alpha R_{c}\Bigg[\frac{\big(\frac{R}{R_{c}}\big)^{n}}{1+\big(\frac{R}{R_{c}}\big)^{n}}\Bigg],
\end{eqnarray}
respectively, with $\alpha$, $\beta$ and $n$ being the arbitrary free parameters and $R_{c}$ parametrises the curvature scale. For each model, different analytical constraints on these parameters is discussed in more detail in the papers by \cite{Abebe2013, Nojiri2017, Cardone2012, Starobinsky2007, Tsujikawa2008b, Motohashi2015}. We also used the effective cosmological constant term ($\Lambda \equiv \frac{\beta R_{c}}{2})$ to mimic dark energy, to allow us to solve these realistic models \cite{Starobinsky2007, Odintsov2017}. The reason for not only using the two realistic model, but also using toy models, is to test how the MCMC simulation and the method holds up against models that have disadvantages, such as the first toy model not being supported by observations or even valid for GR when $n\not = 1$ \cite{Clifton2005}\footnote{In this paper using observational constraints, they determined that for $f(R) = R^{1+\delta }$ to be valid in a GR spacetime, $\delta$ is constrained to lie within the range $0\leq \delta <7.2\times 10^{-19}$ \cite{Clifton2005}.}. This will give as another indication on how well the method and MCMC simulation works.

Even though only 4 models are listed, we ended up with 8 different models that we have tested, since we found that except for the first toy model, the models become analytically unsolvable. Therefore, we assumed fixed $n$-values for the second toy model, to which we found four different solvable models. We then tried this approach for the two realistic models and were unsuccessful in this approach. This led us to incorporate a numerical optimization method into the MCMC simulation to find an approximated $H^{2}$ value at a particular $z$-value. Using this method, we were then able to build a solution map for different approximated $H^{2}$-values at different redshift values between $0\leq z\leq z^{\prime}$. Using the solution map, we were then able to numerically integrate over $z$ using the Simpson integration rule. From here on out the MCMC Redsimulation were able to calculate the approximated distance modulus value for each supernova. Due to the resolution of the numerical methods, we found that for the Starobinsky model, 3 of the free parameters, did not effect the outcome of the predicted model. This led us to also try to fit a reduced version of the Starobinsky model. 

A question that may arise at this stage is: How were we able to write the \textit{f(R)}-gravity Friedmann equation (currently a function of the scale factor) into a normalized Friedmann equation form (a function of redshift), while having measurable quantities that we can use as free parameters, as was done for eq. \ref{eq. normalized friedmann equation} (e.g. by using $\Omega_{m} = \frac{\rho_{m}}{3H^{2}_{0}}$). To answer this, we firstly had to rewrite eq. \ref{eq: f(R) Friedmann equation} into a more usable form (\textit{shown in Appendix: \ref{App: A}}), since we did not have a measurable quantity for some on the terms in eq. \ref{eq: f(R) Friedmann equation} (e.g $\dot{R}$ and $\ddot{H}$). After using the definitions of the Hubble parameter, the Ricci scalar, the Deceleration parameter, and the Jerk parameter, namely $H = \frac{\dot{a}}{a}$, $R = 6\big(\dot{H}+2H^{2}\big)$, $q = -\frac{\ddot{a}a}{\dot{a}^{2}}$, and $j = \frac{\dddot{a}a^{2}}{\dot{a}^{3}}$, we were able to rewrite eq. \ref{eq: f(R) Friedmann equation} into the form
\begin{equation}
	H^{2}(t)=\frac{1}{qf^{\prime}}\bigg[\frac{\rho_d }{3}- \frac{f}{6} +6H^{4}\big(2+q-j\big)f^{\prime\prime}\bigg].
\label{eq: f(R) new friedman}
\end{equation}

We were now able to substitute the different \textit{f(R)}-gravity models into eq. \ref{eq: f(R) new friedman} and then solve for $H^{2}(t)$. However, this Friedmann equation, for each specific \textit{f(R)}-gravity model, is still a function of the scale factor and need to switched to a function of redshift. Therefore, we will need to use a parametrisation in terms of redshift for the cosmographic series terms \cite{Capozziello2019a}. We decided to use the parametrisations for these parameters as given in \cite{Cunha2007}. They defined the deceleration parameter as
\begin{equation}
	q_{z} = q_{0}+q_{1}\frac{z}{1+z}\;, 
\label{eq: Deceleration}
\end{equation}
while the jerk parameter was given as a function of the deceleration parameter
\begin{equation}
	j(q) = q(z)\left[2q(z)+1\right]+\frac{q_{1}}{1+z}\;,
\label{eq: Jerk}
\end{equation}
where $q_{0}$ is the current deceleration parameter value and $q_{1}$ is correction. After the insertion of the cosmographic terms, as well as various other changes that were also needed for the $\Lambda$CDM model, the model can then be normalised to find the normalised Friedmann equation, which can then be used by the distance modulus (\textit{shown in Appendix: \ref{App: B}}). Therefore, up to this point we have not used any simplification, just pure substitution of different definitions equations to get it into a measurable form, with the only exception being the arbitrary parametrisation of the cosmographic terms. But these parametrisations are just one set, others can be used but the more complex they become the more free parameters appear in the model. It must be noted that this is a same method as the one presented in \cite{Nunes2017}. They just went the route of finding a free parameter ($b$) to encapsulate all of the free parameters, while we kept all of the different parameters.

\begin{figure}[tbp]
\centering 
\includegraphics[width=.48\textwidth,trim = 20 5 42 18,clip]{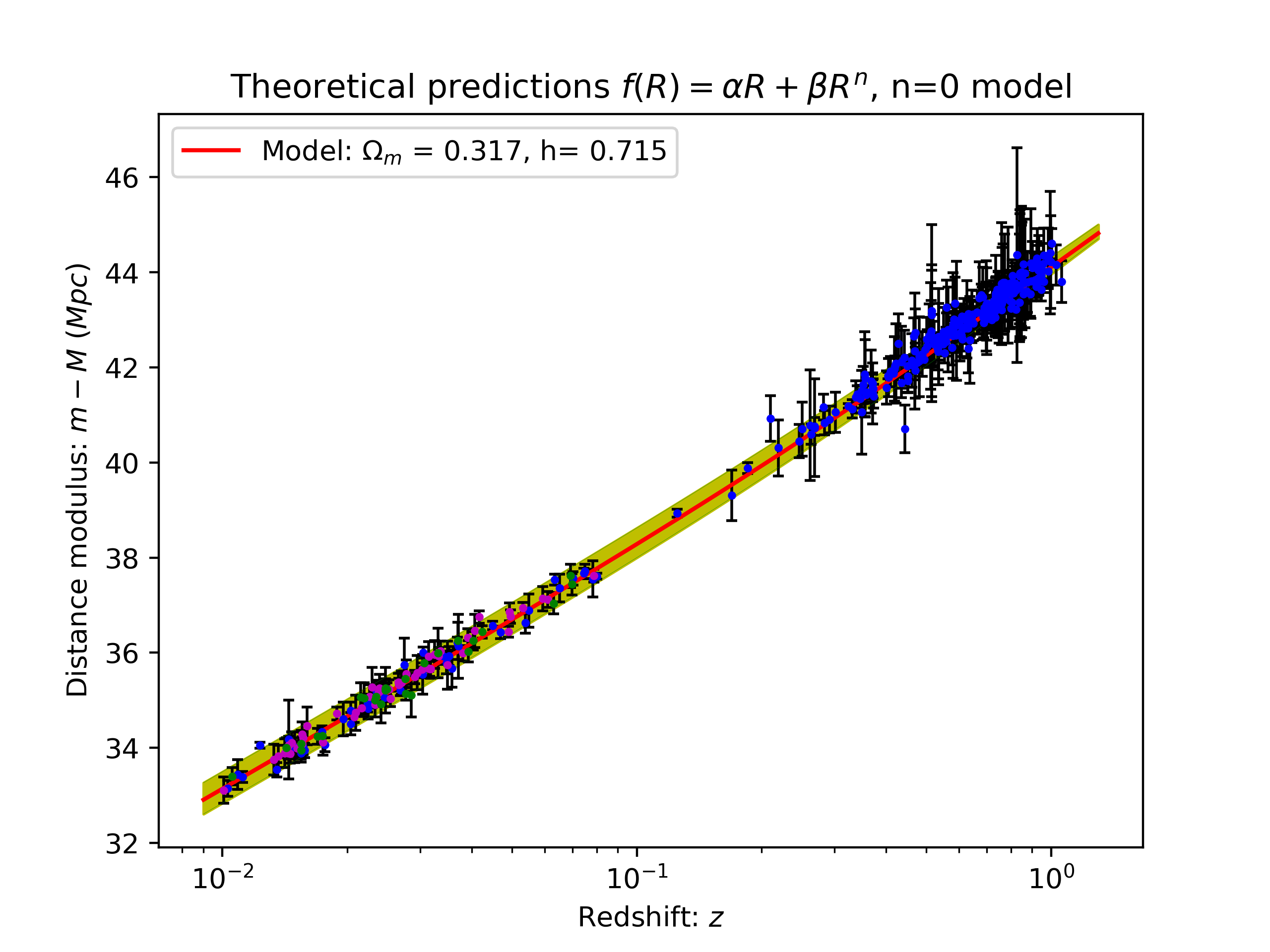}
\hfill
\includegraphics[width=.48\textwidth,trim = 20 5 42 18,clip]{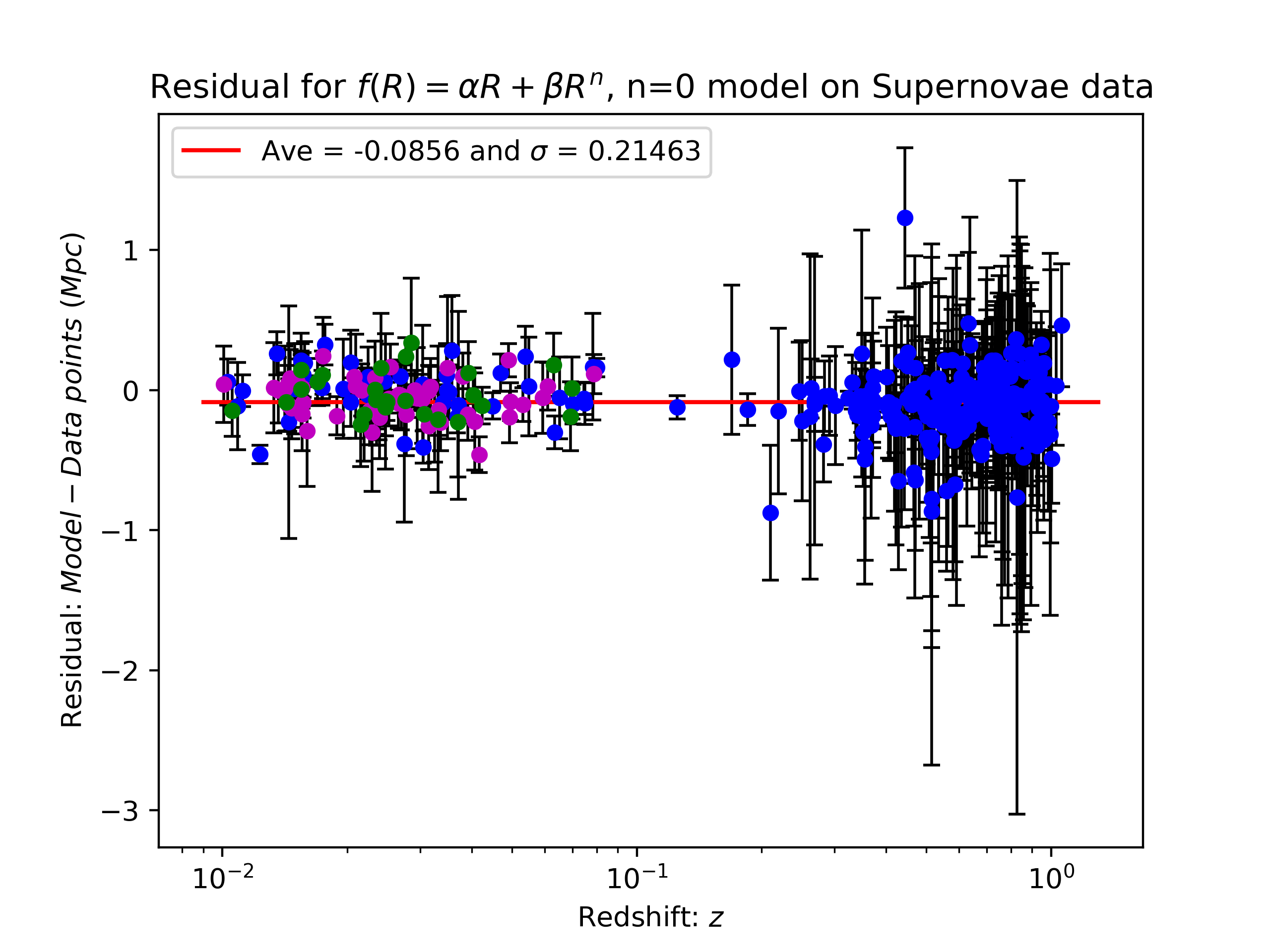}
\caption{The second toy model (with $n=0$) fitted to the Supernovae Type 1A data. With cosmological parameter values calculated by the MCMC simulation as $\Omega_{m}=0.317^{+0.061}_{-0.101}$ (unreliably constrained) and $H_{0} = 71.5^{+6.0}_{-7.2}\frac{km}{s.Mpc}$ (unconstrained), while the arbitrary free parameters were calculated to be $\alpha = 1.202^{+0.397}_{-0.392}$ (constrained) and $\beta  =-5.265^{+1.698}_{-1.315}$ (constrained).}
\label{fig: second toy model n=0 model result graphs}
\end{figure}

We are now able to find the best-fitting function for each of the different \textit{f(R)}-gravity modes, however, due to space limitations we will only present the models that seemed to be able to explain the supernovae data to an extend. Starting in the order that were given above, our first model to show promise is the second toy model, where we assumed $n=0$. Therefore, we have $f(R) =\alpha R+\beta$. The best-fitting model on the supernovae data is shown in figure \ref{fig: second toy model n=0 model result graphs}.

\begin{figure}[tbp]
\centering 
\includegraphics[width=.48\textwidth,trim = 20 5 42 18,clip]{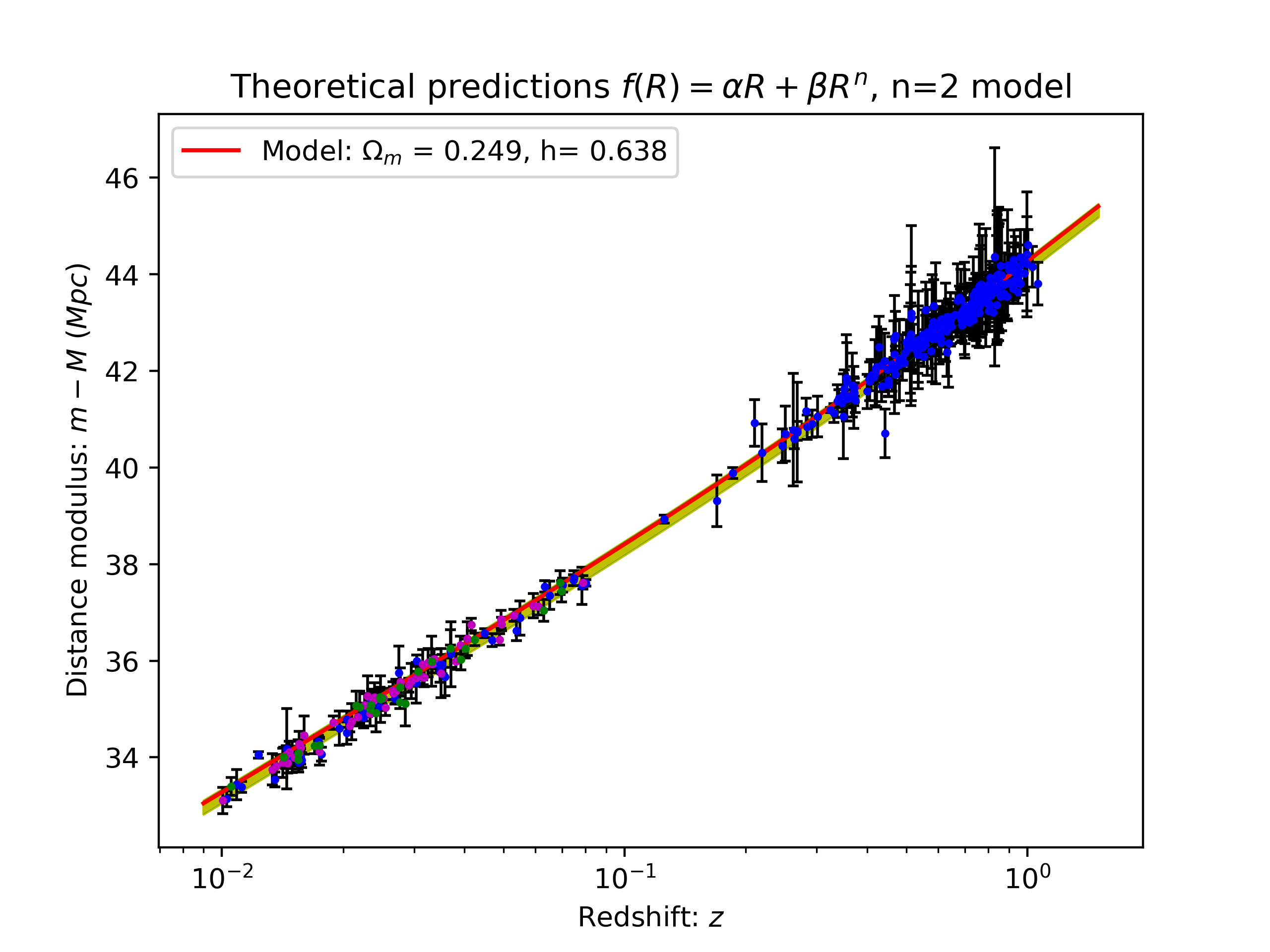}
\hfill
\includegraphics[width=.48\textwidth,trim = 20 5 42 18,clip]{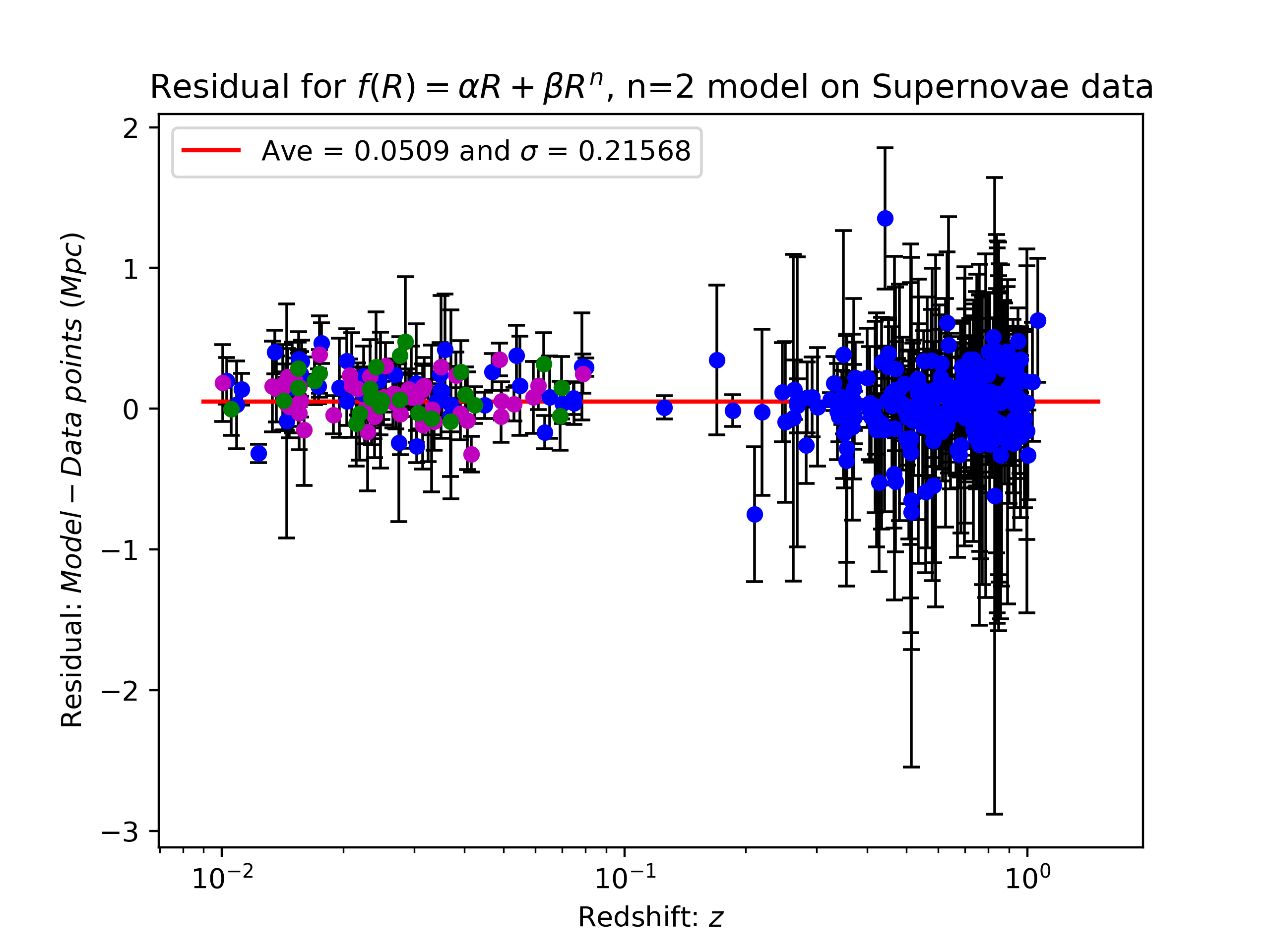}
\caption{The second toy model (with $n=2$ - negative solution) fitted to the Supernovae Type 1A data. With cosmological parameter values calculated by the MCMC simulation as $\Omega_{m}=0.249^{+0.102}_{-0.101}$ (unconstrained), $H_{0} = 63.8^{+4.6}_{-2.7}\frac{km}{s.Mpc}$ (unreliably constrained), $q_{0}=-0.575^{+0.040}_{-0.046}$ (constrained), and $q_{1}=-0.633^{+0.081}_{-0.049}$ (unreliably constrained), while the arbitrary free parameters were calculated to be $\alpha = 19.642^{+2.967}_{-1.753}$ (unreliably constrained) and $\beta  =0.903^{+0.070}_{-0.107}$ (unreliably constrained).}
\label{fig: second toy model n=2 model result graphs}
\end{figure}

It is interesting that this particular model is able to explain the data, since this model resembles the $\Lambda$CDM model. By this we mean that if $f(R) = R-2\Lambda$ (therefore $\alpha=1$ and $\beta = 2\Lambda$), it would be exactly the same as the $\Lambda$CDM model \cite{Odintsov2017}. An important difference between these two models is the fact that the MCMC simulation was only able to fully constrain the arbitrary free parameters and not the cosmological parameters for this \textit{f(R)}-gravity model, while fully constraining both the cosmological parameter for the $\Lambda$CDM model. We did also determined the cosmological constant for this model, if we were to rewrite this model to resemble the $\Lambda$CDM model and found $\Lambda = 2.190^{+1.011}_{-0.900}$. Since this is almost double the values of the cosmological constant, it just shows us the impact of the free parameters. The second model that were able to explain the supernovae data, is also part of the second toy model group, where we fixed $n=2$. This particular model $f(R) = \alpha R+\beta R^{2}$ is also one of the original models developed by Starobinsky to explain the early time expansion \cite{Abebe2013, Starobinsky1980}. Furthermore, this model obtained a positive and a negative solution. We will be showing the negative solution. The best-fitting model is shown in figure \ref{fig: second toy model n=2 model result graphs}.

\begin{figure}[tbp]
\centering 
\includegraphics[width=.48\textwidth,trim = 20 5 45 14,clip]{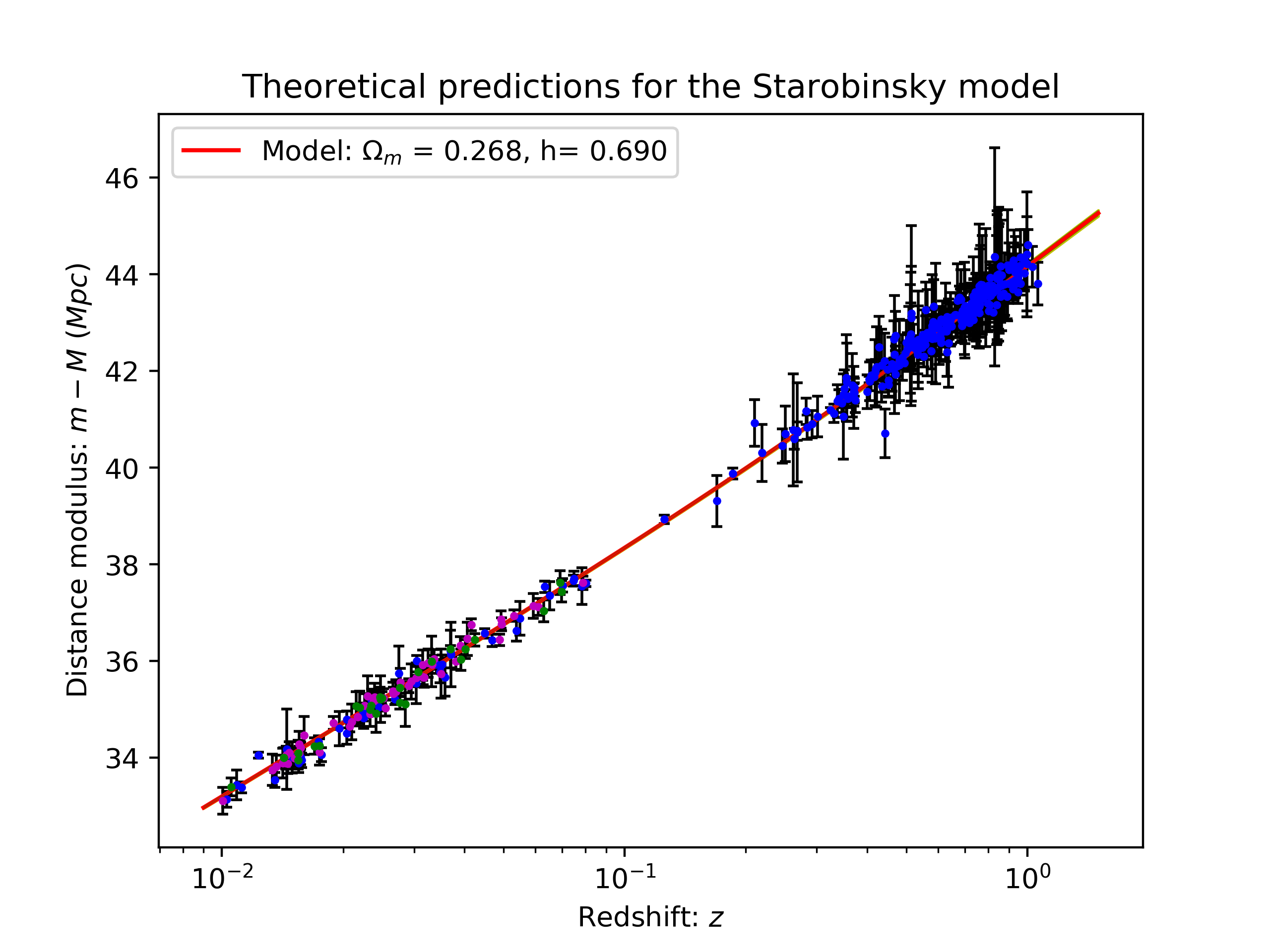}
\hfill
\includegraphics[width=.48\textwidth,trim = 20 5 45 10,clip]{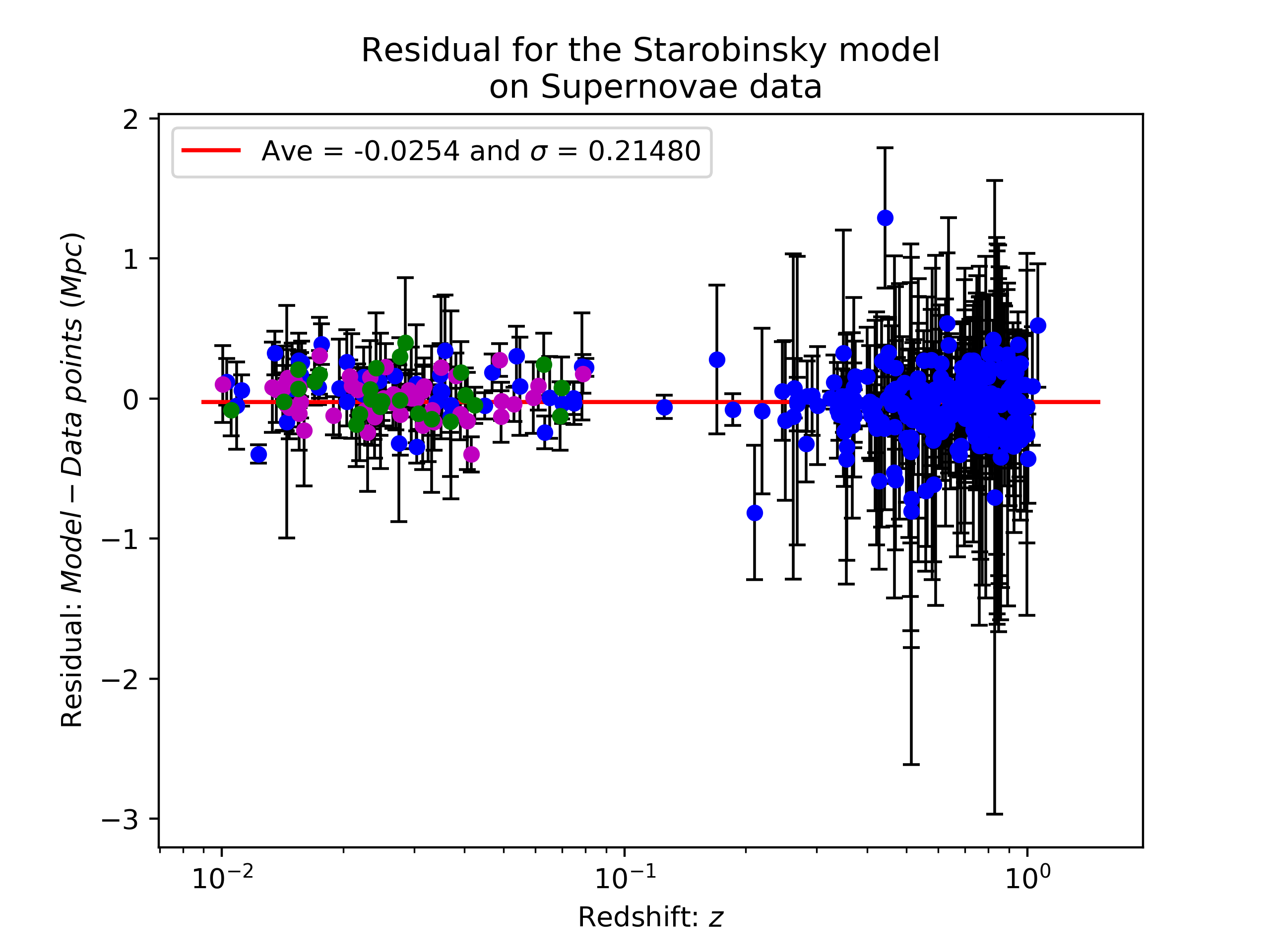}
\caption{The Starobinsky model fitted to the Supernovae Type 1A data. With cosmological parameter values calculated by the MCMC simulation as $\Omega_{m} = 0.268^{+0.027}_{-0.024}$ (constrained), $H_{0} = 69.0^{+0.5}_{-0.5}\frac{km}{s.Mpc}$ (constrained), $q_{0}=-0.512^{+0.328}_{-0.265}$ (unconstrained), and $q_{1}=0.037^{+0.991}_{-1.050}$ (unconstrained), while the arbitrary free parameters were calculated to be $\beta = 5.284^{+3.191}_{-2.981}$ (unconstrained) and $n =4.567^{+3.346}_{-2.899}$ (unconstrained).}
\label{fig: Starobinsky model result graphs}
\end{figure}

Similar to the second toy model where $n=0$, the fixation of $n=2$ also able to explain the data with no over- or under-estimations, although only the deceleration parameter was fully constrained. It is though worth mentioning that this result is somewhat in agreement with the results found by \cite{Guth1981}, where they showed that this model fits the observational data excellently. Even though only the deceleration parameter is the only parameter that is fully constrained. Even though not fully constrained the other parameter results were realistic. This includes the lower than usual Hubble constant (which is still within $1\sigma$ from the CMB results). The last three models that were able to explain the data, was the Starobinsky (with its reduced version) and the Hu-Sawicki model. These were solved using the numerical method. The first potential viable model of these three numerically calculated results is the Starobinsky model, which actually obtained a larger likelihood probability prediction than the $\Lambda$CDM model, as well as being our overall best-fitting \textit{f(R)}-gravity model. The best-fitting Starobinsky model is shown in figure \ref{fig: Starobinsky model result graphs}. 

\begin{figure}[tbp]
\centering 
\includegraphics[width=.48\textwidth,trim = 20 5 45 14,clip]{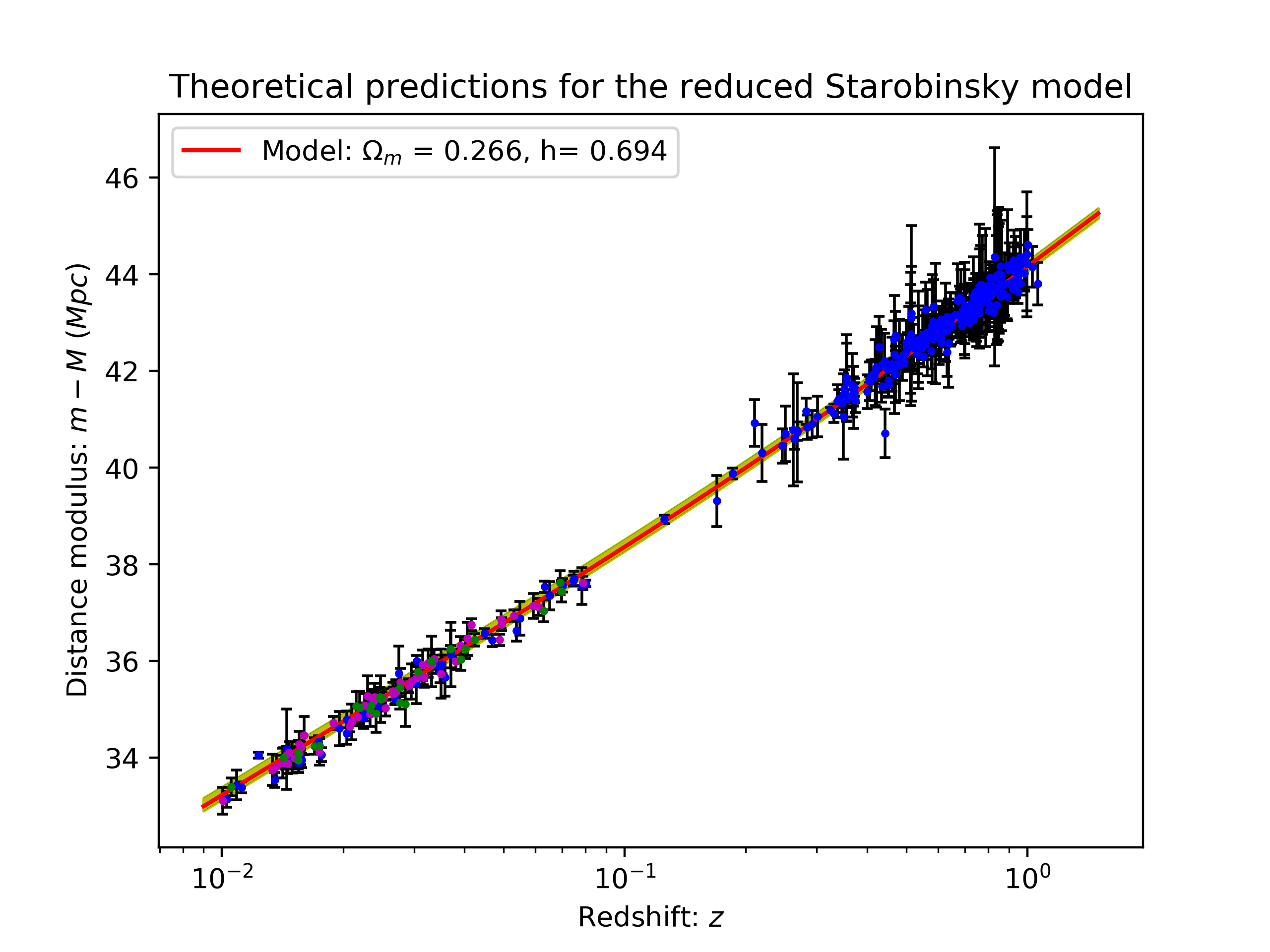}
\hfill
\includegraphics[width=.48\textwidth,trim = 20 5 45 10,clip]{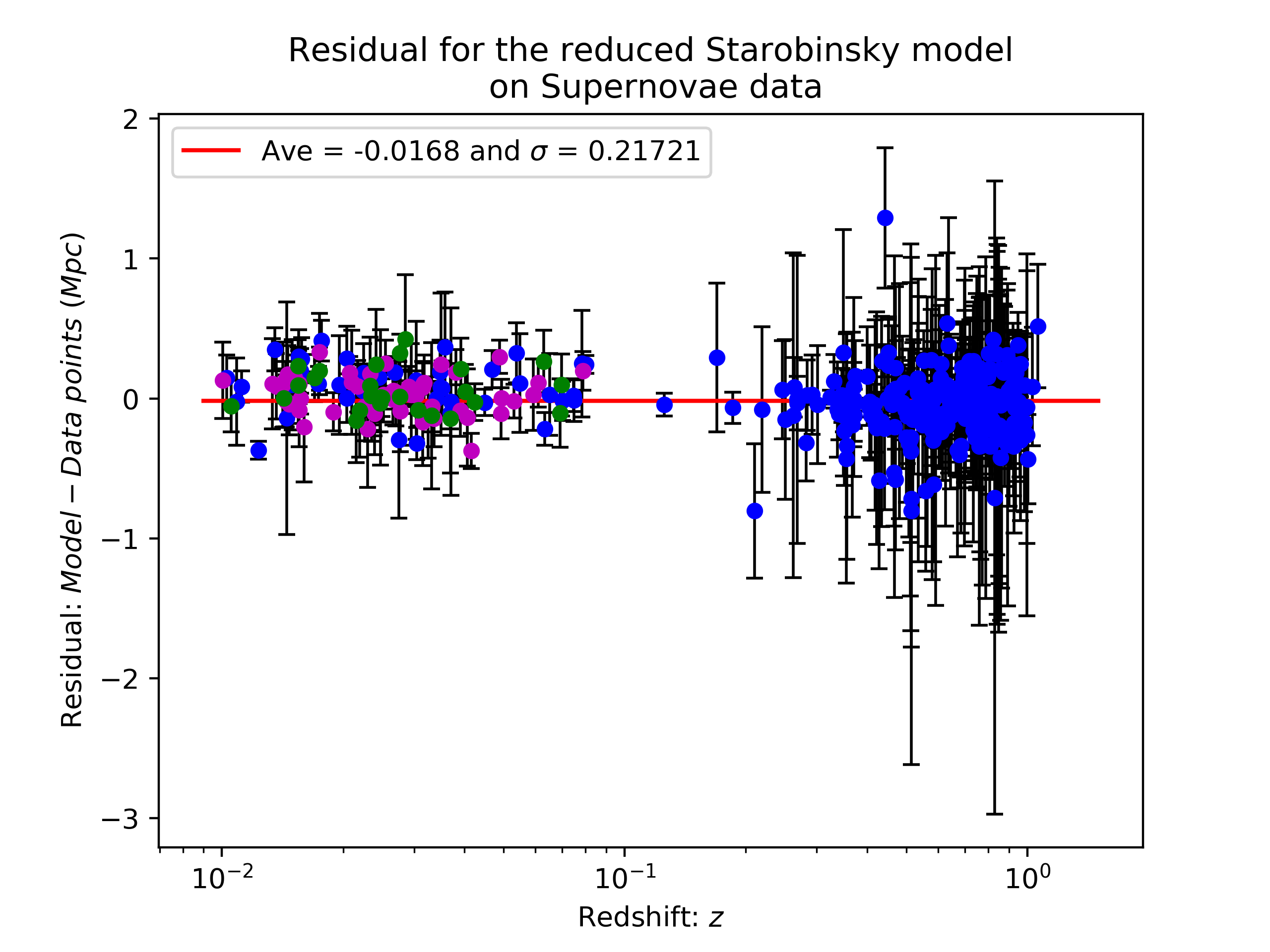}
\caption{The reduced Starobinsky model fitted to the Supernovae Type 1A data. With cosmological parameter values calculated by the MCMC simulation as $\Omega_{m} = 0.266^{+0.026}_{-0.024}$ (constrained), $H_{0} = 69.4^{+1.8}_{-0.6}\frac{km}{s.Mpc}$ (constrained), and $q_{0}=-0.697^{+0.173}_{-0.138}$ (unconstrained).}
\label{fig: reduced Starobinsky model result graphs}
\end{figure}

From figure \ref{fig: Starobinsky model result graphs}, it is clear that the Starobinsky model fits the data with a high precision. Furthermore, we can also assume that this model is quite stable, since the error bars on this model, just like the $\Lambda$CDM model is very small, therefore the MCMC simulation is certain that the predicted best-fit for this model is correct. The only problem faced by this result is the fact that only the cosmological parameters were constrained, while all the remaining free parameters were left unconstrained. Furthermore, due to the model being able to explain the data quite well, we can come to the conclusion that the basic shape of the function is dependent on the cosmological parameters, while the fine-tuning of the function's shape is done by the arbitrary free parameters. However, due to the resolution of the numerical method this fine-tuning is not as effective. This led us to try and find a reduced Starobinsky model with fewer parameters. To reduces this model, we fixed the correctional deceleration parameter to be $q_{1}=0$ (based on the Starobinsky model results). We also fixed $\beta =1$ and $n=1$, after we saw that their error bars are large, but did not translate to large errors in the best-fitting Starobinsky model. Even though this model did not find the accuracy of its counterpart, it was still the third best-fitting model (including the $\Lambda$CDM model) that we found. The results for this reduced Starobinsky model is shown in figure \ref{fig: reduced Starobinsky model result graphs}.

\begin{figure}[tbp]
\centering 
\includegraphics[width=.48\textwidth,trim = 20 5 45 14,clip]{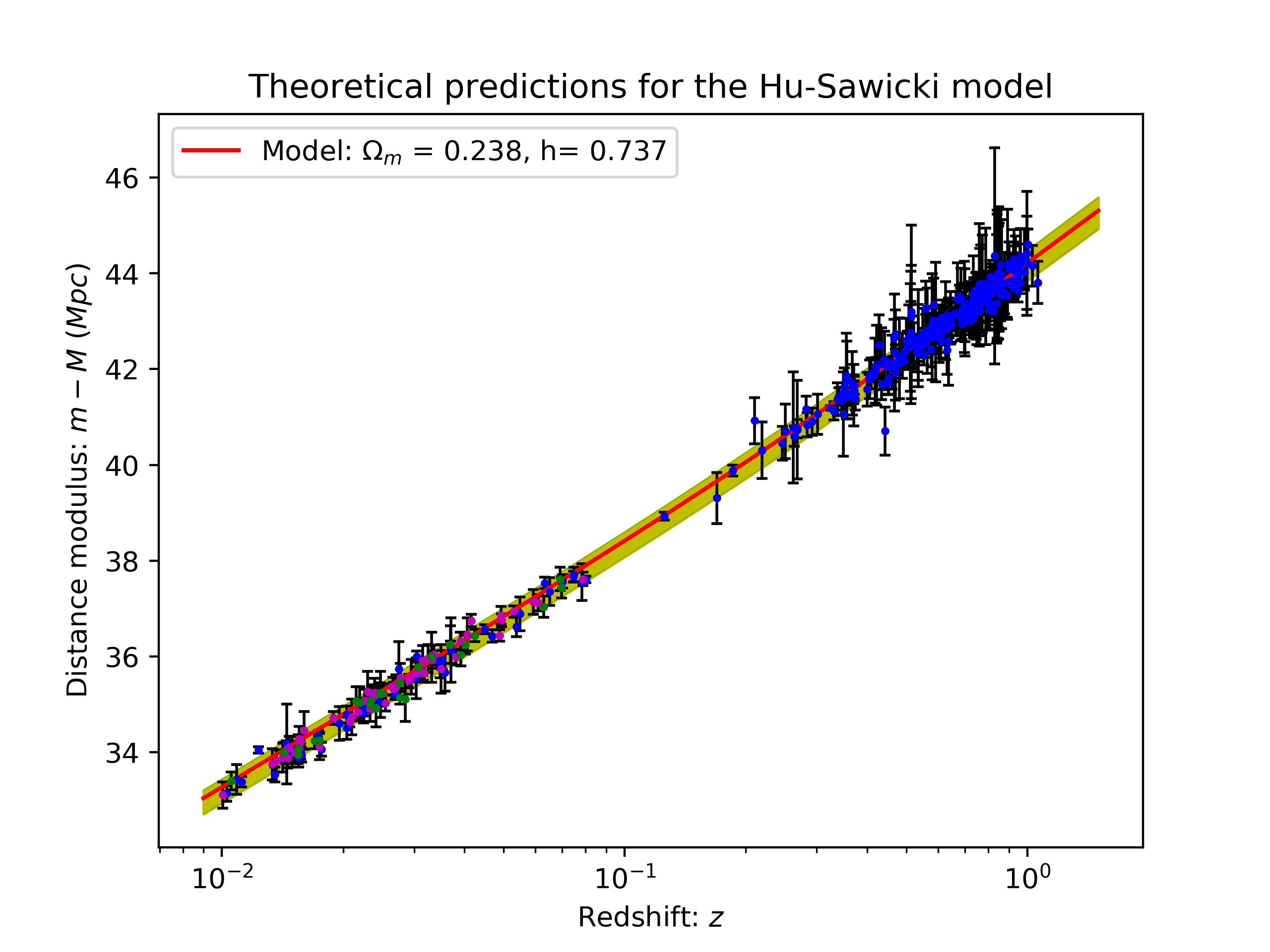}
\hfill
\includegraphics[width=.48\textwidth,trim = 20 5 45 10,clip]{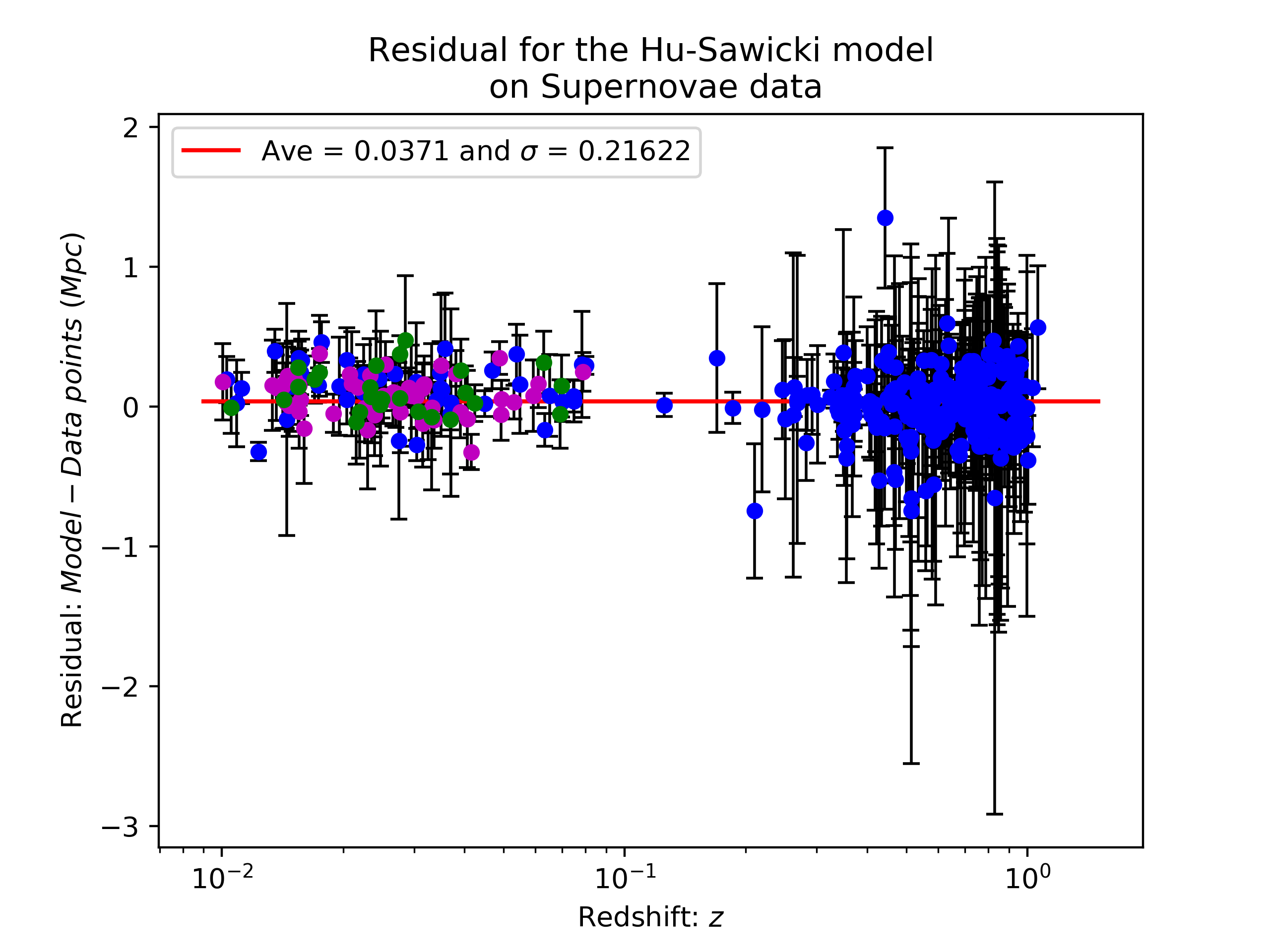}
\caption{The Hu-Sawicki model fitted to the Supernovae Type 1A data. With cosmological parameter values calculated by the MCMC simulation as $\Omega_{m} = 0.238^{+0.043}_{-0.049}$ (constrained), $H_{0} = 73.7^{+9.0}_{-4.6}\frac{km}{s.Mpc}$ (unreliably constrained), $q_{0}=-0.486^{+0.300}_{-0.285}$ (unconstrained), and $q_{1}=-0.036^{+1.018}_{-0.968}$ (unconstrained), while the arbitrary free parameters were calculated to be $\alpha = 5.196^{+2.3224}_{-2.073}$ (constrained), $\beta = 6.923^{+2.120}_{-2.732}$ (unreliably constrained), and $n=2.262^{+0.8Riess200900}_{-0.724}$ (unconstrained).}
\label{fig: Hu-Sawicki model result graphs}
\end{figure}

Due to the fewer free parameters in the reduced Starobinsky model, we can see in figure \ref{fig: reduced Starobinsky model result graphs} that this model is less stable compared to the original model. Therefore, a small change in one of the remaining parameters, can result in a completely different predicted model. It is this fact makes the $\Lambda$CDM model interesting, since it only has 2 free parameters and were still predicting a best-fit model with small errors. We did notice that the deceleration parameter MCMC results were not as uniformed as for the Starobinsky model, suggesting that due to the fewer free parameters the smaller resolution from the numerical method is not as restricting as in the previous case. Lastly, we have the Hu-Sawicki model, which to our surprise did not fair as well or even better than the Starobinsky model, but were still able to explain the supernovae data. The best-fitting Hu-Sawicki model results on the supernovae data is shown in figure \ref{fig: Hu-Sawicki model result graphs}.

From figure \ref{fig: Hu-Sawicki model result graphs}, we can see that even though the Hu-Sawicki model did fit the data, the error region is just as large as the best-fitting function for the $n=0$ second toy model and that was a toy model. This, however, might be an effect of the resolution of the numerical methods, since the Hu-Sawicki model used 7 free parameter, therefore the optimization approximations might have struggled within the MCMC simulation. This is why we kept this model within the group, since it might still be a viable model. For this particular model, we found two constrained parameters, however, only one of the two we a cosmological parameter, namely the matter density distribution parameter. The last three models that we tested, namely the first toy model and the second toy model with $n=\frac{1}{2}$ and $n=1$, obtained best-fitting models that were not able to explain the data.
 
Since we used the two realistic models, namely the Starobinsky and Hu-Sawicki models, we were able to compare them to the results in the research paper by \cite{Nunes2017}, where they used the full JLA dataset, as well as BAO data, cosmic chronometer data and $H_{0}$ observational data, on a state-of-the-art Monte Carlo program, called CLASS, in Python. They found their cosmological parameter values for the Starobinsky model as $\Omega_{m} = 0.269^{+0.050}_{-0.042}$ and $\bar{h} = 0.714^{+0.030}_{-0.028}$, while for the Hu-Sawicki model they found it to be $\Omega_{m} = 0.264^{+0.059}_{-0.055}$ and $\bar{h} = 0.722^{+0.042}_{-0.033}$, respectively. However (as mentioned), in this particular paper they used a singular free parameter ($b$) to encapsulate the remaining free parameters, therefore we were not able to compare our arbitrary free parameters to theirs. This, however, remains a significant result, since we found that even with our small testing dataset, our results are within $1\sigma$ from their results.

\begin{figure}[tbp]
\centering 
\includegraphics[width=.48\textwidth,trim = 20 5 45 10,clip]{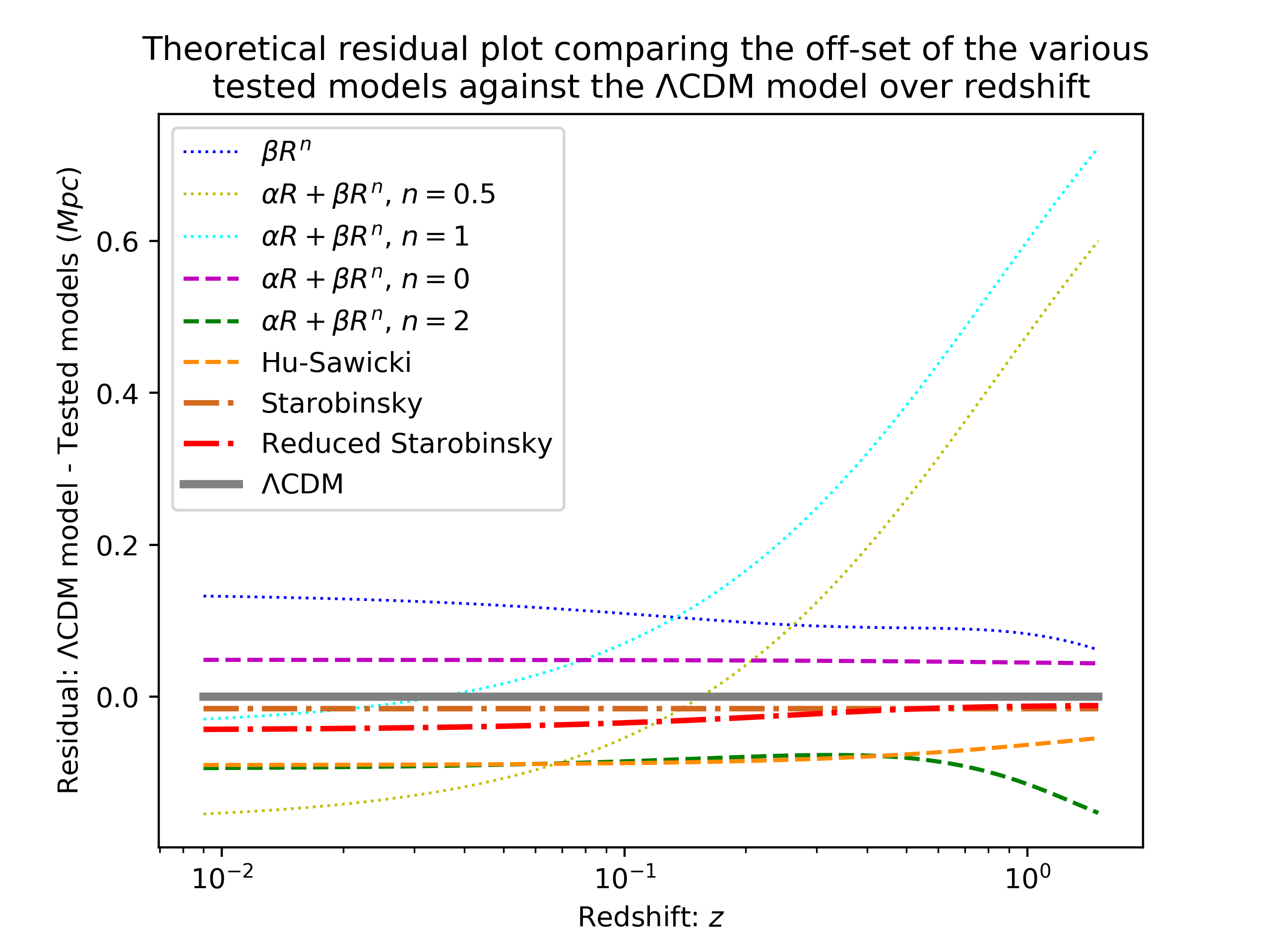}
\caption{Theoretical residuals comparing the different tested models against the $\Lambda$CDM model. The two most succesful models are shown with a \textit{``dashed-dot"} line, while the models that showed promise, were plotted with \textit{``dashed"} lines. The unsuccesful models are shown with \textit{``dotted"} lines.}
\label{fig: Theoretical residuals plot}
\end{figure}

Now that we went through the results of the five best-fitting \textit{f(R)}-gravity models, we can compare them and the three models that were not successful in explaining the data against the $\Lambda$CDM model. To do this we created a theoretical residuals plot between the distance modulus function of the $\Lambda$CDM model and the different \textit{f(R)}-gravity models. This residual plot is shown in figure \ref{fig: Theoretical residuals plot}.

As excepted, the first toy model shows a divergence from the $\Lambda$CDM model for low-redshift due to its incompatibility with GR, with the exception of $n=1$. For the second toy model, however, we have different outcomes. Firstly, we see that for $n=\frac{1}{2}$ and $n=1$, they are not even close to matching the $\Lambda$CDM model in the matter-dominated epoch, however, they do converge rapidly onto the $\Lambda$CDM model, especially for the $n=1$ model, that joins up with the Starobinsky model for low-redshift ($z<0.04$). Therefore, in low-redshift, the second toy model for $n=1$ can explain the data. This is not unexpected, since this from is in a strange way Einstein gravity. It is though still rejected statistically due to its large over-estimation on the distance modulus for the intermediated-redshift supernovae. As for the $n=\frac{1}{2}$ model, it does converge on the $\Lambda$CDM model, but then over-correct and end up being the model that has the largest under-estimation for the distance modulus of the supernovae data in comparison with the $\Lambda$CDM model.

For $n=0$, as noted above, it is simply the $\Lambda$CDM model in terms of arbitrary free parameters, and we do see that is is almost perfectly parallel to the $\Lambda$CDM model, eventhough it is over-estimating the particular distance modulus with less the $0.1$ Mpc relative to the $\Lambda$CDM model. For $n=2$, which is the simplified form of the Starobinsky inflationary model \cite{Starobinsky1980}, this model converges to the $\Lambda$CDM model for the intermediate redshift, and the entering the dark energy epoch it follows the trend set by the $\Lambda$CDM model, as expected, since this model was developed for an accelerating universe. It under-estimates the distance modulus with about $\sim 0.1$ Mpc for the late-time acceleration wit regards to the $\Lambda$CDM model. The Hu-Sawicki model follows the same trend as the second toy model for $n=2$, with the exception that it diverges away from the $\Lambda$CDM model while in the matter-dominated epoch, and then almost matches the simplified Starobinsky inflationary model for the dark-energy epoch. 

This leaves us then with the the two Starobinsky models. It is clear from figure \ref{fig: Theoretical residuals plot}, that this two models, matches the $\Lambda$CDM model the most closely from all of the different \textit{f(R)}-gravity models. Both these models start almost identically in the matter-dominated epoch, however, at the transition phase, the reduced Starobinsky model diverges a bit from the original Starobinsky and $\Lambda$CDM model. This can be due to the limitations we added manually to the Starobinsky model to simplify it without any physical reason, only to see how the model will be affected by reduction in the number of free parameters. However, it is still the third best-fitting model, including the $\Lambda$CDM model. Both of the Starobinsky models under-estimates the distance modulus of the $\Lambda$CDM model with less than $0.05$ Mpc.

\subsection{Statistical analysis}
\begin{table*}
\caption{The best fit for each test model, including the $\Lambda$CMD model. The models are listed in the order from the largest likelihood function value $\mathcal{L}(\hat{\theta}|data)$ to the smallest likelihood of being viable. The reduced $\chi^{2}$-values are given as an indication of the goodness of fit for a particular mod4el. The AIC and BIC values are shown, as well as the $\Delta IC$ for each information criterion. The $\Lambda$CDM model is chosen as the ``true model".}
\label{tab:statistical analysis} 
\begin{tabular}{llllllll}
\hline\noalign{\smallskip}
\textbf{Model} & \textbf{$\mathcal{L}(\hat{\theta}|data)$} & $\chi ^{2}$ & Red. $\chi ^{2}$& \textbf{$AIC$} &\textbf{$|\Delta AIC|$}& \textbf{$BIC$} &\textbf{$|\Delta BIC|$}\\
\noalign{\smallskip}\hline\noalign{\smallskip}
Starobinsky & -120.7052 & 241.4105 & 0.6839 & 253.4105 & 7.9939 & 276.7104 &23.5272\\
			$\Lambda$CDM & -120.7083 & 241.4166 & 0.6762 & 245.4166 & 0 & 253.1832 & 0\\
Starobinsky red. & -122.4442 & 244.8885 & 0.6879 & 250.8885 & 5.4719 & 262.5385 & 9.3553\\
			$\alpha R +\beta$& -131.2518 & 262.5037 & 0.7394 & 270.5037 & 25.0871 & 286.0370 &32.8538\\
			Hu-Sawicki& -140.1668 & 280.3336 & 0.7964 & 294.3336 & 48.9170 & 321.5169 &68.3336\\
			$\alpha R +\beta R^{2}$& -155.0369 & 310.0738 & 0.8784 & 322.0738 & 76.6572 & 345.3737 &92.1905\\
			$\beta R^{n}$& -175.0105 & 350.0211 & 0.9916 & 362.0211 & 116.6045 & 385.3210 & 132.1378\\
			$\alpha R +\beta \sqrt{R}$& -347.0748 & 694.1496 & 1.9664 & 706.1496 & 460.7330 & 729.4496 &476.2664\\
			$\alpha R +\beta R$& -488.3049 & 976.6099 & 2.7510 & 984.6099 & 739.1933 & 1000.1432 &746.9600\\
\noalign{\smallskip}\hline
\end{tabular}
\end{table*}

We are now able to do a statistical analysis on all the different \textit{f(R)}-gravity models, to firstly find their goodness of fit, and secondly to determine whether they are statistically viable alternative models to explain the expansion of the Universe. Using all of the criteria from section \ref{sec:stats}, we can set-up table \ref{tab:statistical analysis}.

From table \ref{tab:statistical analysis}, we see that the two Starobinsky models obtained likelihood function values that are close of even better than the $\Lambda$CDM model, and only obtained a percentage deviation on the goodness of fit of $\approx 1.14\%$ and $\approx 1.73\%$ respectively. However, based on the goodness of fit from the reduced $\chi^{2}$, the $\Lambda$CDM model still fits the supernovae data better than the two Starobinsky models. The other 3 models that were shown in the previous section, can still be considered good fits, since their $\chi^{2}$-values are still relatively close to the $\Lambda$CDM model, with the weakest fit (second toy model with $n=2$) between these 5 models having an $\approx 30\%$ deviation on the ``true model's" goodness of fit. It must be noted that by weakest fit, we do not say that the model does not explain the data, it is just not the best. For example, it was statistically rejected, but its $\chi^{2}$-value on its own is still an excellent fit. It is also evident is the residuals plot figure \ref{fig: second toy model n=2 model result graphs}, where its average over-estimation of the distance modulus compared to the supernovae is $\bar{x}_{res} = 0.0509$ Mpc, which is very small compared to the distances to these supernovae. Therefore, this is still in agreement with the finding of \cite{Guth1981}. It just shows you that there are models that do explain the data better. For the last three models, this percentage deviation, based on the goodness of fit, increases exponentially. 

From the criteria selection, only the two Starobinsky models were deemed viable, with both obtaining a category 2 status for the AIC: ``less support w.r.t. `true model'". However, only the reduced Starobinsky model obtained the category 2 status for the BIC, with the rest all being statistically rejected, even though some were able to fit the data.

Furthermore, we found that the models that obtained constrained parameters, tended to fare better than the models that we left unconstrained. In particular, the five best-fitting models, including the $\Lambda$CDM model all obtained two constrained parameter, while the next best three only obtained one constrained parameter each and the remaining model (not fitting the data at all) did not constrain any free parameters. We also noticed that the models that constrained that cosmological parameters fared better than the models that only constrained the arbitrary free parameters, with the only exception being for the second toy model with $n=0$. This model performed better than the Hu-Sawicki model, even though one of the 2 constrained parameters the Hu-Sawicki model obtained, is the matter-density distribution. However, this might be related to the fact that this particular toy model is in essence the $\Lambda$CDM model, just in terms of \textit{f(R)} gravity. We can, from this knowledge, make the conclusion that the cosmological parameters control the shape of the function while the arbitrary free parameters are used to fine-tune the function to fit the data with a higher precision.

We have now obtained a few different models (five to be exact) through testing whether or not they might be viable alternative models, with the Starobinsky and Hu-Sawicki models obtaining cosmological parameter values that are within $1\sigma$ from the results found in \cite{Nunes2017}. Using different techniques such as increasing our JLA dataset to the full version to improve our statistics, or using other datasets as seen in the research papers of \cite{D'Agostino2019}, or even trying to reduce the number of free parameters like we have done with the reduced Starobinsky model can be done in future work to constrain this group of potential viable \textit{f(R)}-gravity models.

\section{Conclusions}
In this work, we looked at how GR can be used to explain the expansion of the Universe through the usage of the Friedmann equations. This particular set of Friedmann equations, called the $\Lambda$CDM model, had to include the dark energy term to explain the late-time acceleration of the expansion. We then discussed how this model introduces problems due to an early-time acceleration, as well as posing the dark energy problem since it is an unknown pressure force. We then discussed possible alternative modifications to the GR model, which are able to explain the accelerated late-time expansion of the Universe with the exclusion of dark energy. One of these alternative theories is called \textit{f(R)}-gravity.

Following the \textit{f(R)}-gravity model's theory, we looked at how we will be able to find a best-fitting model for different \textit{f(R)} models. This led us to develop a MCMC simulation to fit the distance modulus for each \textit{f(R)} model to Supernovae Type 1A data and find the cosmology parameters ($\Omega_{m}$ and $\bar{h}$). We used the $\Lambda$CDM model to determine whether or not the MCMC simulation was correctly set-up. We also used the $\Lambda$CDM as a ``true model" to compare the \textit{f(R)}-gravity models to it. 

By comparing, firstly, just the residuals of the various tested \textit{f(R)}-gravity models to the $\Lambda$CDM model, we already noticed that the models that tended to be more realistic, such as the original Starobinsky (and its reduced version) and the Hu-Sawicki model had a similar trend than the $\Lambda$CDM model's distance modulus. We also saw that the particular models based on the second toy model, that had a connection to realistic models, such as to the $\Lambda$CDM model and the Starobinsky inflationary model, has similar distance modulus functions as the $\Lambda$CDM model, albeit over-or under-estimating it a bit, it also follow the $\Lambda$CDM model's distance modulus trend. While the first toy model continues diverging away from the $\Lambda$CDM model in the dark energy epoch and the other two toy models have very large over-estimations (up to at least $0.5$ Mpc) for the matter dominated epoch.

Statistically, we found the same five different \textit{f(R)}-gravity models that were able to explain the data. In fact the Starobinsky model obtained a larger likelihood of occurring than the $\Lambda$CDM model, however had a slightly worse goodness of fit, with a deviation of $\approx 1.14\%$ w.r.t. to the $\Lambda$CDM model. Therefore, Starobinsky model was only given a category 2 on the Jeffery's scale for the AIC selection, while being statistically rejected by the BIC selection. The reduced Starobinsky had a smaller likelihood of occurring, and a slightly worse fit with a $\approx 1.73\%$ deviation w.r.t. the $\Lambda$CDM model. This model though was the only model to receive a category 2 status on both the AIC and BIC selections. Therefore, its the only model that fits the data and have some statistical significance. The other three models were able to fit to the data, but were statistically rejected. 

By comparing the residuals between the data and the tested models, theoretical residuals between the $\Lambda$CDM model and the tested models, as well as doing a statistical analysis on these models, we found insights into how these \textit{f(R)}-gravity models compare numerically, not only to the $\Lambda$CDM model, but also how they themselves explain the data. Even though we knew from the beginning that only the realistic models are worth investigating, by testing models that had disadvantages, we were able to test whether the method and the MCMC simulation that we used were successful. Since this method was able to show that these models does not explain the data as expected, we can argue that this method does indeed work. Therefore, the models that the MCMC simulation gave as potential models to explain the data has more validity.

In terms of constraining these five model's parameter values, we found that the models that obtained more constrained parameters, especially the cosmological parameters, tended to fit the data better that the models with fewer constrained parameters. Therefore, if we are to use a more efficient computer software in the future, where we can constrain all the parameters on different datasets, we will be able to constrain these potential viable models with a higher accuracy. However, it is worth noting that we were able to compare the Starobinsky and Hu-Sawicki models with results from more advance studies and we still found our cosmological parameter values to be within $1\sigma$ from their results. Therefore, we will need to use the different datasets and a more efficient program just to fine-tune constrain our tested models. 

The last 3 models that we investigated were not able to explain the data and were subsequently statistically rejected. Therefore, in future work it will not be necessary to work with them.

\begin{acknowledgements}
Renier Hough acknowledge funding through a National Astrophysical and Space Science Program (NASSP) and National Research Foundation (NRF) scholarship (Grant number 117230). Amare Abebe and Stefan Ferreira acknowledge that this work is based on the research support in part by the NRF (with grant numbers 109257/112131 and 109253 respectively). We also acknowledge the help received from the Centre for Space Research at the North-West University.
\end{acknowledgements}

\paragraph{Masters dissertation:} 4
The work presented in this article is based on the findings in the Masters dissertation of Renier Hough \cite{Hough2019a}. Furthermore, an early results conference proceedings based on this work was submitted \cite{Hough2019b}. The MCMC simulation developed in the Masters dissertation, was also used in a group project that were also published in a conference proceedings \cite{Swart2019}. The MCMC codes and the dataset we used in this project can be found on \url{https://drive.google.com/drive/folders/1ag87ouKHzWmuCTBPqsrfLLdRrDmsG4Bi?usp=sharing}.



\begin{appendices}
\section{Finding a usable form for the \textit{f(R)}-gravity Friedmann equation}\label{App: A}
Starting with eq. \ref{eq: f(R) Friedmann equation}, and substituting the expressions for the Ricci scalar and its first-order derivative $\dot{R} = 6(\ddot{H}+4H\dot{H})$, into this Friedmann equation, we find\footnote{We left out the dependencies, such as $f\rightarrow f(R)$, due to the long nature of these equations, but we did take them into account in the mathematical manipulations that we used to determine $h(z)$ for each model.}
\begin{equation}
	H^{2}(t) = \frac{\rho _{m} }{3f^{\prime}} - \frac{\kappa}{a^{2}}+\frac{1}{6}\bigg[6\big(\dot{H}+2H^{2}\big) - \frac{f}{f^{\prime}}\bigg] - H\Big[6(\ddot{H}+4H\dot{H})\Big]\frac{f^{\prime\prime}}{f^{\prime}}.
\label{f(R) friedmann usable form 1}
\end{equation}
We can then use the following known expressions for the Hubble parameter, which is given as:
\begin{equation}
\begin{split}
	\bullet\quad &\dot{H} = -H^{2}(1+q),\\
	\bullet\quad &\dot{H} = \frac{\ddot{a}}{a}-\frac{\dot{a}^{2}}{a^{2}},\\
	\bullet\quad &\ddot{H} = \frac{\dddot{a}}{a}-3\frac{\dot{a}\ddot{a}}{a^{2}}+2\frac{\dot{a}^{3}}{a^{3}}.\\
\end{split}
\end{equation}
By substituting these Hubble parameter definitions into eq. \ref{f(R) friedmann usable form 1} and simplifying, we obtain
\begin{equation}
	H^{2}(t)=\frac{\rho_{m}}{3qf^{\prime}} - \frac{\kappa}{qa^{2}}- \frac{f}{6qf^{\prime}}-\frac{6}{q}\bigg(\frac{\dot{a}\dddot{a}}{a^{2}}+\frac{\dot{a}^{2}\ddot{a}}{a^{3}}-2H^{4}\bigg)\frac{f^{\prime\prime}}{f^{\prime}}.
\label{f(R) friedmann usable form 2}
\end{equation}
By using the definition of the Jerk parameter, we can mathematically manipulate the first term in the bracket to find $\frac{\dot{a}\dddot{a}}{a^{2}}=jH^{4}$, while using the deceleration parameter on the second term in the bracket to find $\frac{\dot{a}^{2}\ddot{a}}{a^{3}} = - qH^{4}$. We can then substitute these expression into eq. \ref{f(R) friedmann usable form 2}, as well as simplifying the equation to obtain
\begin{equation}
\begin{split}
	H^{2}(t)=\frac{1}{qf^{\prime}}\bigg[\frac{\rho_{m} }{3}- \frac{\kappa f^{\prime}}{a^{2}}- \frac{f}{6} +6H^{4}\big(2+q-j\big)f^{\prime\prime}\bigg].
\end{split}
\end{equation}
Since we assumed a flat universe ($\Omega_{k}=0$) for simplicity, we know that $\kappa = 0$. Therefore, we obtain
\begin{equation}
\begin{split}
	H^{2}(t)=\frac{1}{qf^{\prime}}\bigg[\frac{\rho_{m} }{3}- \frac{f}{6} +6H^{4}\big(2+q-j\big)f^{\prime\prime}\bigg].
\end{split}
\end{equation}
\section{Finding $h(z)$ for the Starobinsky inflationary toy model}\label{App: B}
In this section we will present the mathematical steps necessary to find \textit{f(R)}-gravity Friedmann equation in terms of redshift ($H(z)$), as well as the normalized Friedmann equation for the second toy model (with $n=2$), namely the inflationary Starobinsky model. Starting of with just the general second toy model
\begin{equation}
	f(R) = \alpha R+\beta R^{n}.
\end{equation}
You then need to re-parametrise this function, to obtain a dimensionless equation. This is given in \cite{Abebe2013} as
\begin{equation}
	f(R) = \alpha R+H_{0}^{2(1-n)}\beta R^{n},
\label{eq: f(R) =aR+bRn}
\end{equation}
We then need to find the first and second order derivatives, in accordance with eq. \ref{eq: f(R) new friedman}. This we obtain as
\begin{equation}
\begin{split}
	\bullet\quad &f^{\prime}(R) = \alpha+n H^{2(1-n)}_{0}\beta R^{n-1},\\
	\bullet\quad &f^{\prime\prime}(R) = n(n-1) H^{2(1-n)}_{0}\beta R^{n-2}.\\
\end{split}
\label{eq: f(R) =aR+bRn derivatives}
\end{equation}
We can then substitute eqs. \ref{eq: f(R) =aR+bRn} and \ref{eq: f(R) =aR+bRn derivatives} into eq. \ref{eq: f(R) new friedman} and obtain
\begin{equation}
\begin{split}
	&\Scale[1.1]{H^{2} = \frac{\rho_{m}}{3q\Big(\alpha+n H^{2(1-n)}_{0}\beta R^{n-1}\Big)}-\frac{\alpha R+H^{2(1-n)}_{0}\beta R^{n}}{6q\Big(\alpha+n H^{2(1-n)}_{0}\beta R^{n-1}\Big)}}\\
	&\Scale[1.1]{+\frac{6H^{4}\big(2+q-j)}{q}\Bigg[\frac{n(n-1) H^{2(1-n)}_{0}\beta R^{n-2}}{\alpha+n H^{2(1-n)}_{0}\beta R^{n-1}}\Bigg]}.
\end{split}
\label{eq: f(R)=aR+bRn 1}
\end{equation}
We can now substitute the definition equation for the Ricci scalar into eq. \ref{eq: f(R)=aR+bRn 1}, to obtain
\begin{equation}
\begin{split}
	&\Scale[1.15]{H^{2} = \frac{\rho_{m}}{3q\Big(\alpha+n H_{0}^{2(1-n)}\beta 6^{n-1}H^{2(n-1)}(1-q)^{n-1}\Big)}}\\
	&\Scale[1.15]{-\frac{\alpha 6H^{2}(1-q)+H_{0}^{2(1-n)}\beta 6^{n}H^{2n}(1-q)^{n}}{6q\Big(\alpha+n H_{0}^{2(1-n)}\beta 6^{n-1}H^{2(n-1)}(1-q)^{n-1}\Big)}}\\
	&\Scale[1.15]{+\frac{6H^{4}(2+q-j)}{q}\Bigg[\frac{n(n-1) H_{0}^{2(1-n)}\beta 6^{n-2}H^{2(n-2)}(1-q)^{n-2}}{\Big(\alpha+n H_{0}^{2(1-n)}\beta 6^{n-1}H^{2(n-1)}(1-q)^{n-1}\Big)}\Bigg]}.
\end{split}
\end{equation}
We can now solve for $H^{2}(t)$ using the Maple mathematics program and obtain
\begin{equation}
\begin{split}
	&\Scale[0.80]{H^{2}=RootOF\Bigg(6(q-1)\bigg[\beta H_{0}^{2}\Big[\alpha H^{2}(j-2 - q)n^{2}}\\
	&\Scale[0.80]{ + \Big(\frac{\rho_{m}(q-1)}{3}-H^{2}(q^{2} + j - q - 3)\alpha\Big)n + \alpha H^{2}(q-1)^{2}\Big]\Big(-6H^{2}(q-1)\Big)^{n}}\\
	&\Scale[0.80]{+6H^{2}\alpha(q-1)^{2}\big(\alpha H^{2} - \frac{\rho_{m}}{3}\big)H_{0}^{2n}\bigg]-\Big[(j-2 - q)n^{2} + (2q-q^{2} - j+ 2)n}\\
	&\Scale[0.80]{ + (q-1)^{2}\Big]\beta^{2}H_{0}^{2(2-n)}n\Big(-6H^{2}(q-1)\Big)^{2n}\Bigg)}.\\
\end{split}
\label{f(R)=aR+bRn 2}
\end{equation}
Since this equation is not analytically solvable, we insert $n=2$ into eq. \ref{f(R)=aR+bRn 2}. We then simplify using Maple to find
\begin{equation}
\begin{split}
	&-5184H^{2}\Bigg[\beta\bigg(j -\frac{1}{2}q^{2} - q - \frac{3}{2}\bigg)H^{4} + \frac{\alpha H^{2}H^{2}_{0}}{12} - \frac{\rho_{m}H^{2}_{0}}{36}\Bigg]\\
	&\times (q-1)^{3}\Bigg[\beta (q-1)H^{2} - \frac{\alpha H^{2}_{0}}{12}\Bigg]=0.
\end{split}
\end{equation}
When solving for $H^{2}$, we obtain 4 different solutions given by
\begin{equation}
\begin{split}
	\bullet\quad H^{2}&=0,\quad\textrm{or}\quad\bullet\quad H^{2} = \frac{\alpha H^{2}_{0}}{12\beta (q-1)},\\
	\bullet\quad H^{2}&=\pm\frac{H_{0}\sqrt{\alpha^{2}H_{0}^{2} - 8\beta\rho_{m}\big( q^{2} - 2j + 2q +3\big)}}{12\beta\big(2j-q^{2}- 2q - 3\big)}\\
	&-\frac{H_{0}^{2}\alpha}{12\beta\big(2j-q^{2}- 2q - 3\big)} .\\
\end{split}
\end{equation}
The first solution is a stationary universe, while the second is only a function of the free parameters, which does not help us in being able to compare the cosmological parameters of the various model. Using the 3rd and the 4th solutions, we can determine the normalised Hubble parameter as a function of time. This we find to be
\begin{equation}
\begin{split}
	h(t) &= \Bigg[\pm\frac{\sqrt{\alpha^{2} - 24\beta\Omega_{m}\big( q^{2} - 2j + 2q +3\big)}}{12\beta\big(2j-q^{2}- 2q - 3\big)}\\
	&-\frac{\alpha}{12\beta\big(2j-q^{2}- 2q - 3\big)}\Bigg]^{\frac{1}{2}}
\end{split}.
\label{eq: normalized friedmann equation}
\end{equation}
To change the dependency of time to redshift, we need to use $\Omega_{m}(t)=\Omega_{m} (1=z)^{3}$, as well as the parametrised cosmographic series terms as defined in eqs. \ref{eq: Deceleration} and \ref{eq: Jerk}. By substituting these terms into eq. \ref{eq: normalized friedmann equation}, to obtain
\begin{equation}
\begin{split}
	h(z) &= \Bigg[\pm\frac{\sqrt{\alpha^{2} - 24\beta\Omega_{m}(1+z)^{3}\big( q^{2}(z) - 2j(z) + 2q(z) +3\big)}}{12\beta\big(2j(z)-q^{2}(z)- 2q(z) - 3\big)}\\
	&-\frac{\alpha}{12\beta\big(2j(z)-q^{2}(z)- 2q(z) - 3\big)}\Bigg]^{\frac{1}{2}}.
\end{split}
\label{growth_rate aR+bRn_n2}
\end{equation}
We showed the MCMC simulation's result for the negative solution.
\end{appendices}

\end{document}